\documentclass[fleqn,10pt]{wlscirep}

\usepackage[utf8]{inputenc}
\usepackage[T1]{fontenc}
\usepackage{mathtools}
\usepackage{times}
\usepackage{lineno}
\usepackage{xcolor}
\usepackage{xspace}
\usepackage{hyperref}
\usepackage{comment}
\usepackage{geometry}
\usepackage{graphicx}
\usepackage{amsfonts}
\usepackage{amsmath}
\usepackage{booktabs}

\title{The rise and fall of WallStreetBets: social roles and opinion leaders across the GameStop saga}

\author[1,2,*]{Anna Mancini}
\author[1,2,3]{Antonio Desiderio}
\author[4,2]{Giovanni Palermo}
\author[5,6]{Riccardo {Di Clemente}}
\author[1,2,+]{Giulio Cimini}
\affil[1]{Physics Department and INFN, University of Rome Tor Vergata, 00133 Rome (Italy)}
\affil[2]{Centro Ricerche Enrico Fermi, 00184 Rome (Italy)}
\affil[3]{DTU Compute, Technical University of Denmark, 2800 Copenhagen (Denmark)}
\affil[4]{Physics Department, Sapienza University of Rome, 00185 Rome (Italy)}
\affil[5]{Complex Connections Lab, Network Science Institute, Northeastern University, London E1W1LP (United Kingdom)}
\affil[6]{ISI Foundation, 10126 Turin, (Italy)}
\affil[*]{anna.mancini@cref.it}
\affil[+]{giulio.cimini@roma2.infn.it}
\date{}

\begin{abstract}
    Nowadays human interactions largely take place on social networks, with online users' behavior often falling into a few general typologies or ``social roles''. Among these, opinion leaders are of crucial importance as they have the ability to spread an idea or opinion on a large scale across the network, with possible tangible consequences in the real world. In this work we extract and characterize the different social roles of users within the Reddit WallStreetBets community, around the time of the  GameStop short squeeze of January 2021 -- when a handful of committed users led the whole community to engage in a large and risky financial operation. 
    We identify the profiles of both average users and of relevant outliers, including opinion leaders, using an iterative, semi-supervised classification algorithm, which allows us to discern the characteristics needed to play a particular social role. The key features of opinion leaders are large risky investments and constant updates on a single stock, which allowed them to attract a large following and, in the case of GameStop, ignite the interest of the community. Finally, we observe a substantial change in the behavior and attitude of users after the short squeeze event: no new opinion leaders are found and the community becomes less focused on investments. Overall, this work sheds light on the users' roles and dynamics that led to the GameStop short squeeze, while also suggesting why WallStreetBets no longer wielded such large influence on financial markets, in the aftermath of this event.
\end{abstract}

\begin{document}
\flushbottom
\maketitle
\thispagestyle{empty}
\section*{Introduction}

Nowadays the quest of understanding human behavior has entered a highly computational and quantitative era \cite{manifesto_css}, thanks to the availability of data from social media and new technologies that capture interactions at the level of individual users over extended periods of time \cite{lazer_css}. 

We can collect e-mail data \cite{email_net}, credit card data \cite{Di_Clemente_2018}, records of phone calls \cite{phone_data} and all kinds of digital traces an individual leaves behind when roaming in the online world \cite{browsing_his}. 
The availability of a huge amount of microscopic data allows us to deepen our understanding of human behavior by employing powerful scientific methods, such as those from statistical physics, applied to a different phenomenon: people \cite{RevModPhys.81.591,cimini_statphys}.
Remarkably, social media and web technologies not only provide us with such richness of data, they bring human interaction itself to a different dimension. Interactions between millions of individuals are now possible, people can be virtually connected to everybody and everything with the touch of a finger, accessing and sharing information almost instantly with the rest of the world \cite{social_media_review}. These factors give rise to new emergent phenomena: polarization \cite{polarization_bail}, diffusion of fake news \cite{fakenews}, echo chambers \cite{echochamber}, protests and revolutions \cite{chileprotests,arabspring}, mass retail investment \cite{semenova_bullruns}, just to name a few. 

Among the various social networks that have been analysed in the literature, Reddit represents an ideal playground \cite{recurringpatterns,reddit_opioid}, since users are mostly anonymous, thus data are not affected by any privacy issue. 
Moreover, Reddit is organised into forums (subreddits) devoted to specific topics, like politics \cite{politosphere}, cryptocurrencies \cite{reddit_crypto} and many more. 
The interest in Reddit has risen to par with other platforms like Facebook and Twitter due to its recent increase in popularity caused by the GameStop (GME) short squeeze. 
In January 2021, users of the Reddit community WallStreetBets (WSB) coordinated in a mass effort to buy and hold GameStop shares, which drove the price to a 30-fold growth in less than a month, causing large losses for all hedge funds who were profiting by short-selling GME \cite{yolo_capitalism,sel-induced_reddit}.

Such an unprecedented event attracted the attention of the academic community, resulting in a stream of literature that analyzed different aspects of the episode.
For instance, Lucchini et al. \cite{lucchini_committed} studied the commitment of WSB users towards the short squeeze, finding how a small fraction of users may have triggered all others to follow through buying GME shares. 
Zheng et al. \cite{social_dynamics_GME} investigated the social dynamics of the event by looking at dynamic interaction networks, topic modeling and sentiment analysis. 
Kim et al. \cite{social_informedness} examined trading patterns of individual investors, finding that social sentiment enhanced the collective behavior. 
Haq et al. \cite{users_wsb_gme} analyzed the different behaviors of "old" users versus the behavior of those who joined following the success of the short squeeze. 
Mancini et al. \cite{sel-induced_reddit} looked for the early warnings of the event and proposed an opinion dynamic model to qualitatively reproduce the sudden emergence of consensus in the community.
However, we still do not have a clear picture of the individual behaviors of users and the social roles they played in the community. Shedding light on this aspect is essential to understand how a handful of individuals managed to have such a large following, as to bring the collective action on GME to success. 
In this work we aim at filling this gap, by developing a framework to classify WSB users and assess what allows them to earn a social role recognized by others. 

The concept of \emph{social role}, defined as a typical behavior associated with a social position, is well known in the social sciences literature \cite{role-theory}.
Some studies have already tackled the issue of extracting and studying these roles on different social networks.

For example, on Wikipedia pages Wesler et al \cite{social_roles_wiki} found that it is possible to identify four key social roles in the community of editors by analyzing the editing history of users. While the online questionnaire put toghether by Brandtzaeg \& Heim \cite{users_typology} informed us on the typology of Norwegian social network users based on their online habits. Moreover, the study carried out by Recuero et al. \cite{roles_polarized_recuero} exploits network topology to identify Twitter user roles and their influence in political discussions by focusing mainly on those with high degree, while Amato et al. \cite{centrality_amato} proposed an approach based on hypergraphs to identify user roles, introducing a new centrality measure used to detect lurkers (users who read but do not actively participate in an online community). Refer to the survey by Jin et al. \cite{user_roles_survey} for a review of the main topics of literature regarding user behavior, from the analyses of interactions on a social graph to the identification of malicious conduct.
However, these works either rely on ad-hoc techniques (such as questionnaires) and/or are based on a predetermined set of user roles to be detected, whose existence is then verified on the data available.
To overcome these limitations, we develop a framework where user roles emerge "naturally" through a semi-supervised procedure which can be tuned for various contexts, starting from a set of individual features that reflect user activity and the quality of the content created. 
In particular, we detect the main social roles of "average" users and their temporal dynamics by informing a DBSCAN \cite{dbscan} algorithm with the selected features. Furthermore, by recursive application of the methodology, we can identify a handful of \emph{outliers}: "special" users who, due to a behavior that consistently deviates from the average, possess peculiar features and social statuses.

A user role of special interest is the so called \emph{opinion leader} \cite{two_step_flow}: a hub of the community who is able to directly influence other members. 
Recently, different approaches have been employed to identify these types of users on social media. Sun et al. \cite{multi-features_sun} selected and analyzed multiple features from Weibo users to single out these leaders. Winter and Neubaum \cite{facebook_ol} directly asked Facebook users how influential they consider themselves. 
Riquelme et al. \cite{milestones_ol} defined a new centrality measure based on activity and interest of users over a specific topic to identify opinion leaders on Twitter. 
Haase et al. \cite{finfluencers} studied how financial influencers on Twitter affect or are affected by the overall sentiment towards a financial asset using Granger causality. 
However all these works are limited to social networks where a user's identity is known, and thus their influence may also be a consequence of the individual's pre-existing popularity. The anonymity of Reddit users instead allows for identification of well grounded features that contribute to achieving the status of opinion leader.
In the case of WSB, where the major topic of interest is whether to invest or not in a given stock, opinion leaders can be considered as those who, through their investments, achieve a large following and sway others to invest in their same stock. By triggering large collective investments, opinion leaders can potentially succeed in influencing the financial market.

Through our semi-supervised framework, we are able to identify user roles and in particular the opinion leaders of WSB across the GME saga -- explaining which features allowed these users to achieve their prominent status. 
By analyzing their activity we find that each opinion leader is linked to a particular stock investment, and we are able to determine how the "GME leader" was able to trigger massive consensus, which contributed to the success of the GME short squeeze. Finally, we detect how roles change after the events of January 2021.
Indeed the mediatic notoriety that WSB achieved in the aftermath of the GME short squeeze changed the very nature of the community, making discussions less focused on finance-related topics and more on memes -- suggesting the community may have lost its potential to trigger successful collective retail investments.

\section*{Results}
The subreddit WallStreetBets (WSB) is a community of mostly amateur traders \cite{wsb_positions_ban} who like to discuss high risk investments and post about their gains and losses. WBS users can write new posts and comments under existing posts, resulting in tree-like conversation threads, like the one shown in Figure \ref{fig1}A. The community has specific rules, one of which states that any claim about an investment has to be supported by a visual proof of the latter, i.e. a screenshot, which gives us valuable information about when such an investment was made. Particular members of the community, called \emph{moderators}, ensure that rules are observed by all users. Given this special role that shapes their behavior, identifying who they are brings useful information in our user role analysis, and we do so by looking at previous literature \cite{social_dynamics_GME} and inspecting user's comments (see Supplementary Materials \ref{sm_bots} and \ref{mod_beh}). The time period that we consider comprises all posts and comments written on the community from August 1, 2020 to July 31, 2021 (see Methods for the precise number of comments and posts).

\subsection*{The rise of WallStreetBets}
\begin{figure}[p]
    \centering
    \includegraphics[width=\textwidth]{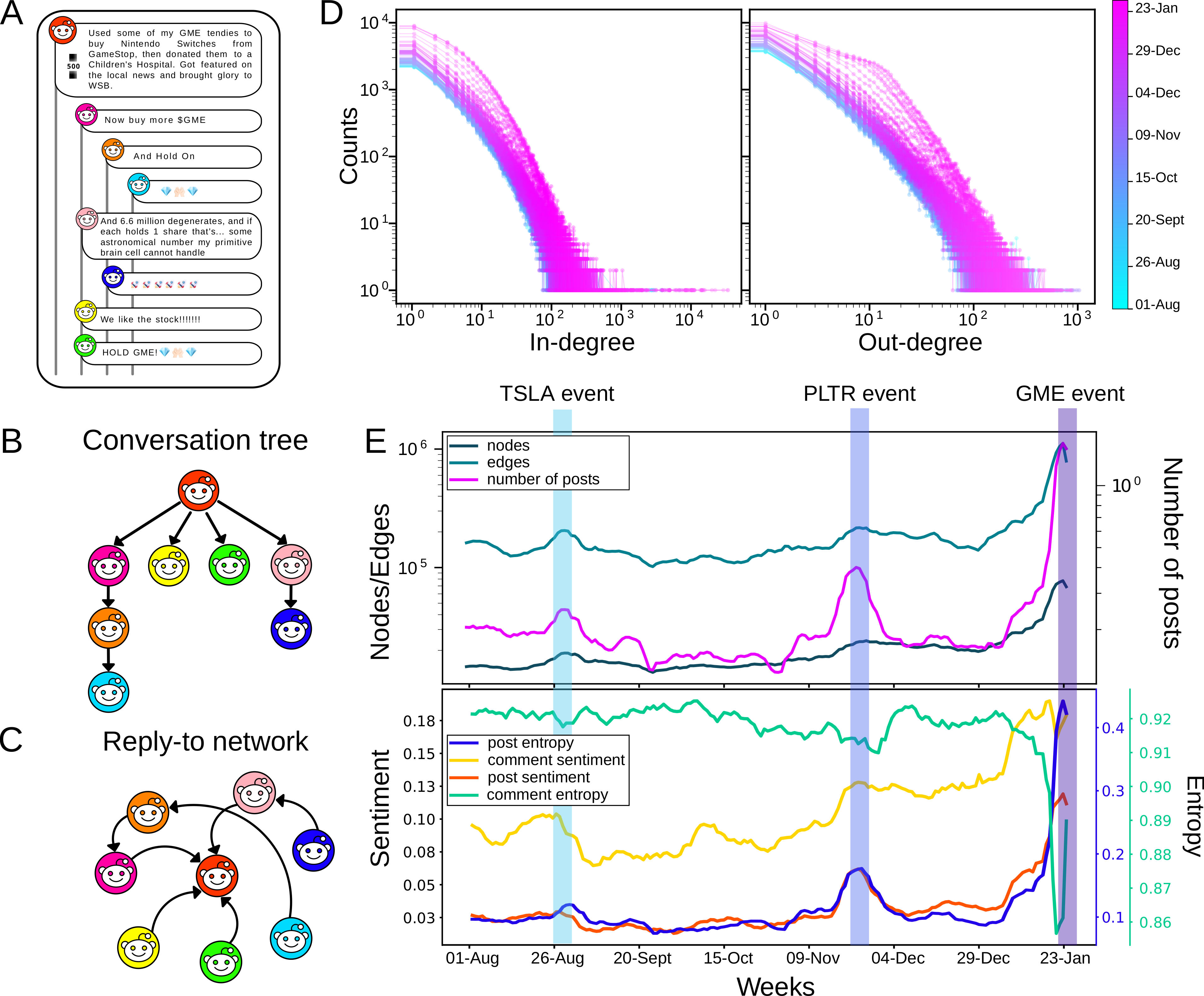}
    \caption{\textbf{Construction of conversation networks and time evolution of user features on WSB.} 
A) Example of a Reddit conversation thread, made up of the post and its comments. B) Conversation tree extracted from the thread. C) User network built from the "reply-to" interactions on the thread. D) In- and out-degree distributions for all weekly user networks, from August 1, 2020 to January 31, 2021. We can see the presence of some highly connected hubs in the in-degree distribution, i.e. users whose comments and posts have received many replies. E) Upper panel: nodes and edges of the user network in time. These peaks in network structure are reflected in those of the number of posts created, which correspond to some engaging events that attracted the attention of the entire community around some stocks. Lower panel: sentiment and entropy of posts and comments in time. Both post and comment sentiment increase when the event occurs. Instead comment entropy diminishes since the text of comments becomes more repetitive, while post entropy peaks because of the presence of longer posts analyzing and explaining the situation.}
    \label{fig1}
\end{figure}

We now focus on the period preceding the short squeeze (August 2020 - January 2021). After a data cleaning procedure to remove bots and inactive users (see Methods) we transform conversations into trees, where users are nodes and a link indicates a direct reply to a post or comment, and then extract interaction networks: 
Figure \ref{fig1}B shows the tree describing the thread of Figure \ref{fig1}A, while Figure \ref{fig1}C is the resulting user "reply-to" network. We obtain weekly networks by considering all "reply-to" actions occurring within a week, and shifting the 7-day moving window by one day each time, until covering the entire timespan of data (see Methods for further information on how these networks are created). 
Figure \ref{fig1}D shows the out- and in-degree distributions of weekly networks, which show a power-law trend. Whereas the in-degree distribution counts grow almost one order of magnitude around January (as a result of the growing number of new users), and highly connected hubs of the community appear; in the out-degree distribution we see a bump forming around 20 as we approach January, meaning that the majority of new users write 20 or less comments per week.

To check how the overall behavior of users is changing in time, we look at the evolution of the number of nodes and edges in the weekly user networks. In the upper panel of Figure \ref{fig1}E we can see how both quantities increase in certain weeks -- when also the number of posts peaks. Inspecting the text of posts we can see how this increase is due to the community's reaction to some event of major interest \cite{recurringpatterns}. We can identify, in temporal order: the stock split of Tesla (TSLA) in August 2020, investor Steve Cohen disclosing a large stake in Palantir (PLTR) in November 2020 \cite{pltr} and the GME frenzy due to the short squeeze in January 2021. Unsurprisingly, posts are the best proxy for the changing interests of the community, and are the primary means though which users share and spread information. Therefore post features play a crucial role in understanding behavioral changes sustained by the community. 

Our main goal is to develop a semi-supervised approach to extract a reliable characterization of user profiles from all the available information we can collect on the way they interact online. Therefore we start by selecting a set of relevant features that capture a user's commenting and posting behavior, the complexity of the comments and posts written and the effect they have on the community. We thus collect features of posts, comments and "reply-to" networks for each user in each week of our dataset (refer to Methods for the complete list and description). In the lower panel of Figure \ref{fig1}E we show the time evolution of a subset of these quantities summarizing the textual complexity of posts and comments written, i.e. the entropy, and the tone that posts and comments convey, i.e. the sentiment, to show how the community as a whole changes in volume, tone and complexity of content posted when particular situations occur. While the sentiment and post entropy all display peaks during the highlighted events, the comment entropy has an opposite trend, since comments tend to become repetitive and plain during an engaging event \cite{recurringpatterns}.

\subsubsection*{User profiles}

In Figure \ref{fig1} we have shown how the community shifts altogether, now we turn to a microscopic level where the focus is on the behavior of single users and how it changes in time. One of the first things we notice by looking at the number of active users and comparing it with the total number of subscribers, is what is known as the \emph{1\% rule}. In agreement with the current literature \cite{social_roles_reddit,1_percent_rule}, we find that only 1\% of users contribute to new content on the community, while 99\% is just lurking. Interestingly, this behavior persists even with the growth of the community around January (see Supplementary Materials \ref{1_rule} for more details). 
Differently from other social networks, Reddit users are mostly anonymous. It has been shown that this peculiarity results in a more dis-inhibited and honest behavior \cite{disinhibition_reddit,veil_anonymity}. This leads to wonder: does anonymity influence the actions and \emph{roles} of users? Or do we find patterns common to other online social networks?

To identify user roles, we represent each user by her feature vector and apply the well-known clustering algorithm DBSCAN \cite{dbscan} to the feature vectors of each week (see Methods for the full explanation of the procedure). We also apply spectral clustering \cite{spectral_clustering} to test the coherence of the obtained results, and achieve an extremely good agreement between the clusters identified by both algorithms (see Supplementary Materials \ref{supp_ARI}).
Figure \ref{fig2}A shows the two-dimensional projection of a sample realization of the clustering algorithm (obtained with the t-SNE algorithm \cite{tsne}) for six different weeks of our dataset. Three major clusters are clearly visible: a large green cluster and two smaller ones, pink and orange, while gray points are unlabelled users classified as noise by DBSCAN.
The time evolution of the clusters outlines how the top three increase in size while some of the smaller ones disappear in correspondence with events that drew the attention of the community, as can be seen from 2-D projections (November 2020: PLTR week, January 2021: GME week). 
In Figure \ref{fig2}B we track the time evolution of the size of the three largest clusters (at least an order of magnitude larger than that of all others). The figure suggests the occurrence of a flow of users from the largest cluster to the other two when the week of the event approaches.

These events make most of the users fall onto some well-defined categories that we are able to describe by analyzing the value of the features in the three biggest clusters. In order of cluster size we identify: \emph{commenters}, users whose post features have very low values; \emph{active users}, with high values of both comment and post features; \emph{posters}, users whose comment features have very low values. The average feature values of the three categories can be seen in Figure \ref{fig2}C for the last week of January (when all three clusters have a high density of points).
Interestingly, the \emph{commenter}, or answer-person, behavior has been found in other social networks \cite{social_roles_online}, as well as in other Reddit communities \cite{social_roles_reddit}, suggesting this as being a very general behavior of online users. In our context we see that, when an event of interest to WSB users occurs, some of the commenters decide to change role and write posts to effectively reach more people, hence becoming active, and many users belonging to a different, smaller, cluster move into one of these three categories, thus reducing the heterogeneity of behavior. Moreover, during the week of the event, active users tend to have a higher textual entropy, both in comments and posts, compared to the other two groups. This might suggest that these users are the ones that facilitate the flow of conversations with a constant input of personal opinions and information.

\begin{figure}[p]
    \centering
    \includegraphics[width=\textwidth]{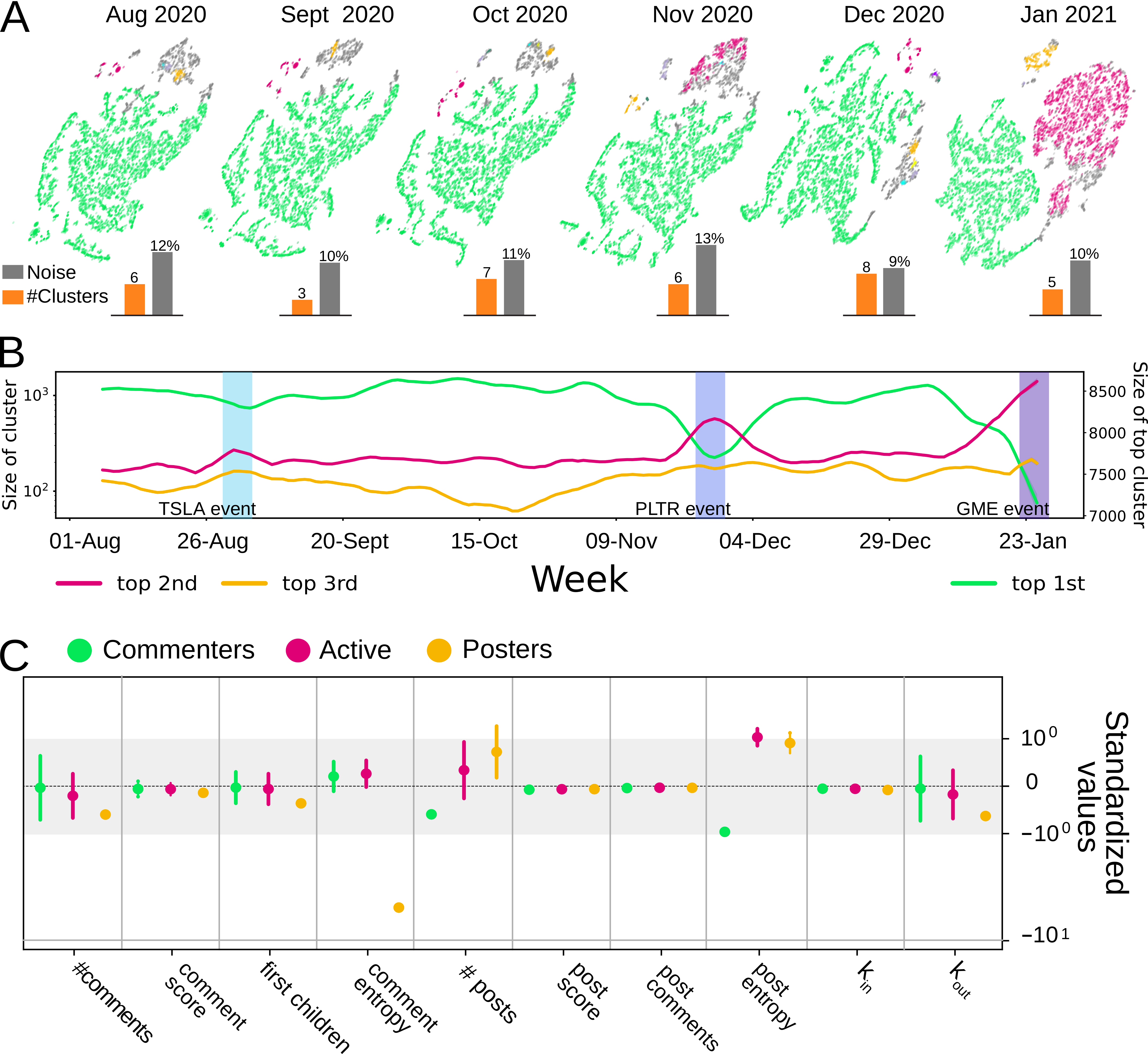}
    \caption{\textbf{User profiles.} 
A) T-SNE projections for the weeks of 24-30 August 2020, 24-30 September 2020, 24-30 October 2020, 24-30 November 2020, 24-30 December 2020 and 24-30 January 2021. The three largest clusters are visible (green: \emph{commenters}, pink: \emph{active}, orange: \emph{posters}), while grey points belong to noise. For each realization we specify the percentage of noisepoints and the number of clusters. The weeks of November and January correspond to engaging events, and we see how clusters become fewer and well separated. B) Time evolution of the sizes of the top three clusters. In the week of the event we have a "flow" of users from the top cluster (commenters) to the other two (active and posters). In fact when an event happens that catches the interest of the community, some of those who normally comment start writing posts and consequently become active or posters. C) Average value of the feature vectors for the three biggest clusters of the week from the 24th to the 30th of January, 2021. We can see how each cluster is characterized by a particular profile of the feature values, which allows us to understand the average behavior of users belonging to such cluster (for the sake of readability we show only selected features on the x axis).}
    \label{fig2}
\end{figure}

\subsubsection*{Outliers} 

We have so far identified the average user behavior on WSB, but we know that there are users that do not fit into any of these average profiles.
To investigate those individuals whose behavior diverges from the average conduct and understand what differentiates them from the rest, we shift our attention to noise points, i.e., users that do not belong to any cluster. Through an iterative clustering-removal procedure (see Methods) we are able to rank users according to how far they lie from the clusters identified by DBSCAN, the largest of which represent the average user profiles. In this way we can select those individuals with a persistent diverging behavior and by labeling each profile depending on its feature attributes we can distinguish different roles -- as reported in Figure \ref{fig3}A. 
In particular we can still identify profiles in line with the commenters, posters and active user roles found in the average user analysis, but with extreme values of their characterising features (for instance, a poster who wrote a tremendous number of posts). 
We also find some new interesting roles: \emph{analyzers}, users who write posts about what is trending on the community and share updates on market news; \emph{bots}, those that were not removed by the initial filter; 
\emph{opinion leaders}, users whose posts and comments have a great following, but who do not comment and post as frequently as some others do. 
In Figure \ref{fig3}B we plot the average number of comments under posts contributed by each user versus the average score of such posts (both quantities are normalized by the total number of daily active users), for all users with more than 10 posts. 
Hence each user is a point in the graph, with size proportional to the number of posts written, while the outlier groups are represented in a different color. 
All three opinion leaders found (light blue) lie in the upper right corner: even if they do not write too many posts, the content they post is always appreciated and generates ample discussion. Triangles instead denote the outliers who are also moderators of the community, and whose profile fits well with the "posters" role. Posters (orange) are found in the right-hand side of the plot and have the largest sizes, as their posts are more frequent and characterized by a high number of comments. Active users (pink) are similar to posters, but their sizes on average tend to be smaller since these users post less. Commenters (green) are mostly in the lower left side of the plot, as they are the ones that post less (as can be seen by their small size) and whose content does not draw much attention. Bots are not found in the plot, since they never write posts. Finally, analyzers (blue) are found in the upper right corner of the plot, highlighting how much their content is appreciated; they have medium sizes since they tend to write posts periodically to update the community.

\subsubsection*{Opinion Leaders}
One of the main issues when trying to identify opinion leaders on social networks is topic limitation. In most cases an opinion leader is an individual with expertise in a particular subject \cite{richmond_2006,grewal_social_identity,opinon_leadership_scale}, and thus does not exert the same authority when speaking of a different matter. Trying to find the topic of interest may be a challenging task \cite{topic-limitation_yang,topic-sensitive_miao}, which could result in biased or erroneous results. This problem is minimized on Reddit since it is structured into communities, each regarding a given subject, and users who comment on these communities are to some degree already knowledgeable on the matter.
In addition, when trying to identify influential users, the starting point is usually to see which ones satisfy some fixed notion of \emph{influence} \cite{influencers_customer_networks}, be it in terms of network position \cite{dg_bodendorf}, feature values \cite{multi-features_sun} and so on. Here instead we want to extrapolate the concept of influence from a behavior we observe, and thus spontaneously deduce which are the actions that allow users to achieve such a position.

\begin{figure}[ht!]
    \centering
    \includegraphics[width=\textwidth]{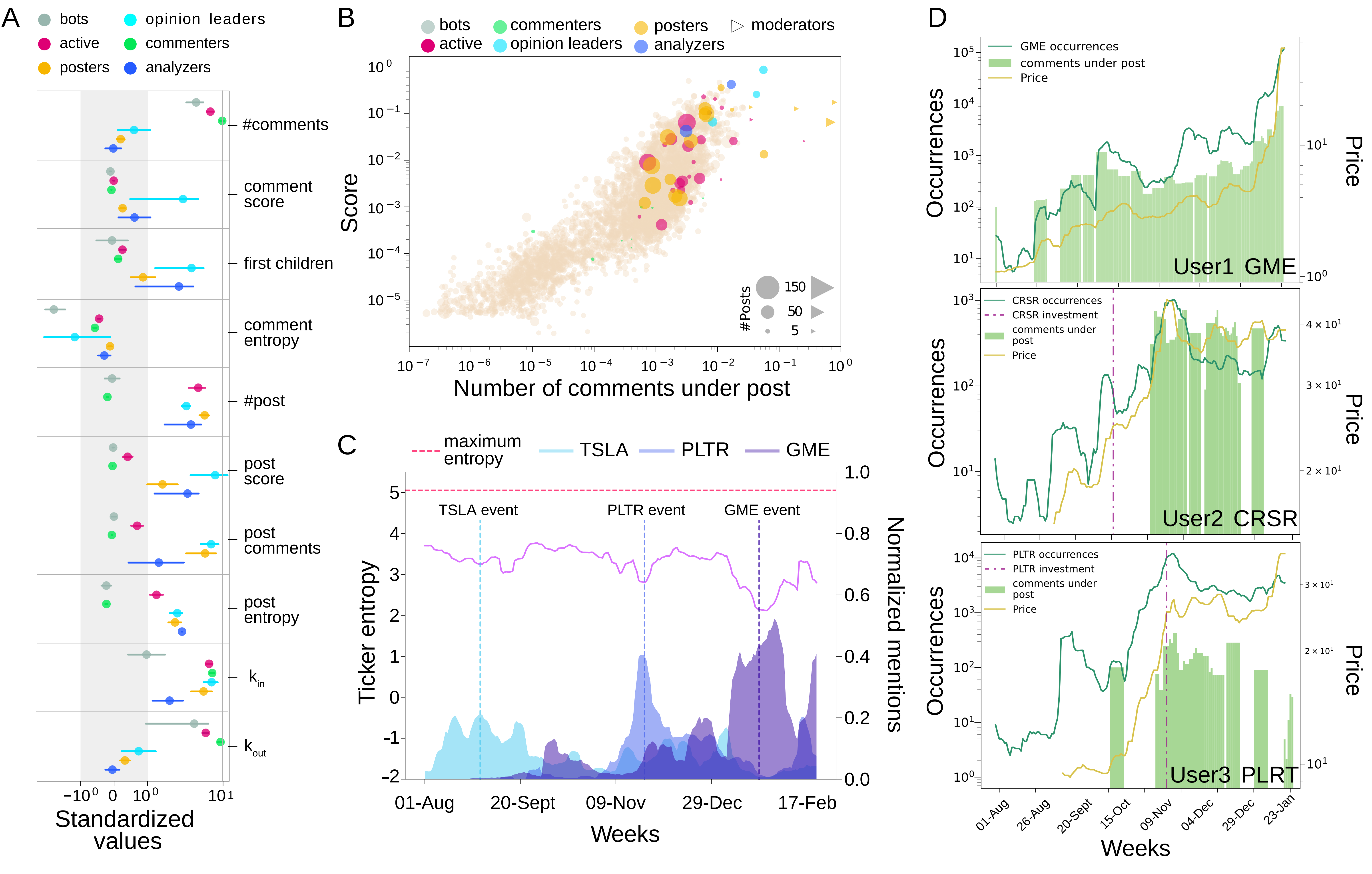}
    \caption{\textbf{Outlier groups.} 
A) The distinct profiles of feature values for the various outlier groups. For the sake of readability not all features are shown on the y axis. B) Average number of comments under posts versus average post score for all users with more than 10 posts, normalized by the total number of daily active users. Each group occupies a well defined region of the plot, with opinion leaders placed in the top right corner and commenters lying in the bottom left one. Each point corresponds to a user with size proportional to the number of posts written. C) Time evolution of the ticker entropy (violet solid line), which decreases each time an engaging event occurs. The minimum is reached on the week of the GameStop short squeeze. The dashed pink line is the maximum value the ticker entropy can assume. The shaded regions correspond to the normalized number of mentions for each of the three engaging tickers. D) For each of the three opinion leaders we report the number of comments under their posts (light green bars), the number of occurrences of their chosen stock ticker on the community (dark green) and the closing price of the stock (yellow). For User2 and User3 we also highlight when the investment was made (purple dashed line), while User1 made the investment in 2019. All quantities grow after the investment, with an increase in the number of comments under their posts right after the increase of the stock price, indicating that these users become more popular after achieving considerable gains.}
    \label{fig3}
\end{figure}

We focus on the posting behavior of the three opinion leaders found (which we refer to as User1, User2 and User3) to identify the attributes that have earned them such visibility. Firstly we noticed that their contributed posts with a large following are all linked to a particular stock investment -- respectively on GME, CRSR (Corsair Gaming Inc) and PLTR. Figure \ref{fig3}D shows several quantities related to such users and their chosen stock: the number of comments under each post (light green bars), the number of mentions of the stock ticker on the community (dark green) and its closing price (yellow). The dashed-dotted line indicates when the investment was made, except in the case of User1, since his investment was made in 2019 (not covered by our data). We can see how after the investment these quantities start to grow, i.e. the stock these users invested in becomes popular among the community, given the rise in number of mentions; the number of comments under their posts becomes of the order of the thousands, thus indicating a boost in their popularity; the price of the chosen stock grows, resulting in considerable gains. By considering all these factors, some common patterns that boosted the leaders' popularity emerge: 
i) Extreme action: all investments are \emph{YOLOs}, i.e. large investment into a not-so-popular stock; such a high risk-high reward approach is highly appreciated by the community, as its name WallStreet\emph{Bets} suggests.
ii) Persistent posting behavior: once the investment is made, all leaders start updating the community on their progress (gains or losses).
iii) Success: the investment has to eventually bring profit to the user, which has happened in all three cases, as can be seen from their investment updates. 

Out of all three opinion leaders, only one was able to completely shift the interest of the community towards his chosen stock. As Figure \ref{fig3}C shows, the number of ticker mentions for each "engaging" stock (the three found in the first section that produced a shift in the community's behavior) normalized by the total number of mentions) reaches a maximum in the week of the corresponding event, while the ticker entropy (see Methods), which is a proxy for how many different stocks are discussed in the community on a given week, decreases in correspondence to the events of interest, but it reaches a minimum in January, when the entire community is engaged in the GME short squeeze. Overall, User1 was exceptional in his behavior, even when compared to the other two opinion leaders. In fact his investment dates back in 2019, and his persistent posting started as a monthly update in that year as well, which became a weekly update in 2020. At the time GME's popularity was declining, so his 50 000 \$ "YOLO" attracted much attention from the community, even if at first many were against him. However, as the first plot of Figure \ref{fig3}D shows, all quantities increase as GME's price starts to rise, consequently resulting in large gains on his side, allowing him to obtain a wide following from August 2020 onwards, to such a degree that moderators created a thread exclusively for GME. The other opinion leaders started investing in the stock as well. All of these events in turn resulted in the full consensus that characterised the GME short squeeze \cite{sel-induced_reddit}.\\
\\

\subsection*{The fall of WallStreetBets}\label{after}

The events of the GME saga put WallStreetBets in the spotlight of the entire world, with a tremendous number of new users who joined from January onwards. Did the community change after such disruptive event and the massive influx of heterogeneous user profiles? Do new opinion leaders emerge?
To find an answer we repeat the same set of analyses described above in the period after the GME short squeeze (February 2021 - July 2021). 

\begin{figure}[ht!]
    \centering
    \includegraphics[width=0.9\textwidth]{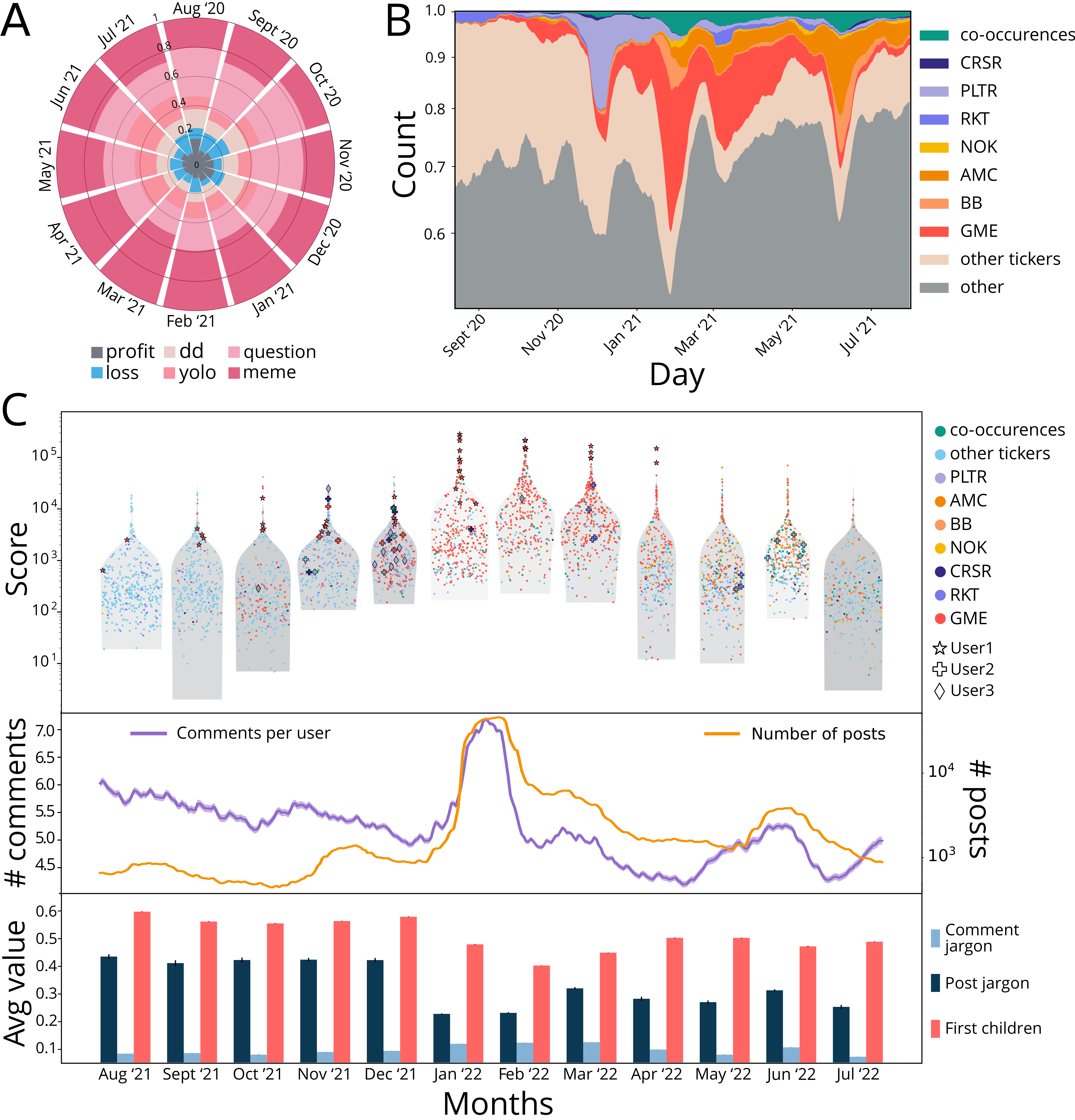}
    \caption{\textbf{The aftermath of the short squeeze.} 
A) Fraction of posts with different labels in the various months before and after the GME event. We observe an increase in meme-labelled posts and a decrease of due diligence ones, highlighting the changing interest of the community. B) Normalized daily count of posts mentioning popular tickers on WSB (GME, BB, AMC, NOK, RKT, PLTR, CRSR) and featuring multiple ticker mentions among them (\emph{co-occurences}), mentioning other tickers (among a set of 157 hand-selected stocks, see Methods \ref{ticker-entropy}) (\emph{other tickers}), and not mentioning any tickers at all (\emph{others}). We see that the number of mentions for a ticker highlights the periods of maximum popularity for the corresponding stock; additionally there is a change in mention patterns after January, resulting in less focused discussions (increase in multiple ticker mentions in a single post) still linked to the effect of the short squeeze (mentions of meme stocks GME, BB, AMC, NOK). We can also see a growth of posts not mentioning any tickers at all. A 14-day moving average is applied to these figures. C) Top: monthly distribution of the top 25 highest ranked posts (by score), with color indicating the ticker mentioned in the post and shaded area proportional to the number posts made in the selected month without any ticker mentions. Posts made by opinion leaders are highlighted and always lie in the upper tail of the distribution. Center: average number of comments per user made each day, with standard deviation of the mean, and number of posts (a 20-day moving average is applied). Bottom: monthly averages of the features: comment jargon, post jargon and first children, with standard deviation of the mean.}
    \label{fig4}
\end{figure}

Amongst all outliers found in this new timespan, only one truly corresponds to the previous description of an opinion leader on WSB: User1 who, even after the hype of the short squeeze died out, was still able to exert a great deal of influence in the community. 
All other popular users we identified lack the most important factor in becoming an influential individual on a community of traders: a successful YOLO investment (or any investment at all, to be precise). Moreover we found that the new subscribers of WSB interact and behave in a different way with respect to the \emph{old} users of WSB \cite{users_wsb_gme}. This can be seen by looking at the \emph{labels} users attach on their posts to characterize the content.
We considered the following labels: \emph{profit}, posts about profits from an investment; \emph{loss}, posts about losses from an investment; \emph{dd} (\emph{due diligence}), lengthy posts with a detailed analysis of a financial situation; \emph{yolo}, posts about large and risky investments; \emph{question}, posts with questions mainly concerning the world of finance; \emph{meme}, posts with memes. Figure \ref{fig4}A shows, for each month, the fraction of posts with a given label. Interestingly we notice that starting from January the \emph{meme} labelled posts increase, while the \emph{dd} ones drop to just a small fraction. 
This change of wording and topics implies that the community's interest has shifted from investing to amusing content: new users post and appreciate memes, while information on actual financial investments gets lost in the multitude of new content created. 

Moreover, posts loose explicit mentions to tickers, as showed by Figure \ref{fig4}B, where we keep track of the ticker content of posts, showing how the fraction of posts involving no tickers (labelled as \emph{other}) grows after January 2021, while posts mentioning tickers diminish (\emph{other tickers}). The only exception to this trend are posts about meme stocks (GME, BB, AMC, NOK), which still make up a relevant fraction of the ticker content created after January. Indeed, in the wake of the enthusiasm of the GME event, users tried to initiate new short squeezes for these stocks, succeeding in their intent with AMC Theatres (AMC) in May 2021 \cite{amc_squeeze}. Moreover we can see how posts mentioning more than one meme stock (\emph{co-occurences}) grow substantially after January, indicating again a less focused financial dialogue.

The top panel of Figure \ref{fig4}C shows the monthly distribution of the daily top 25 posts with highest score. The color highlights the ticker mentioned in the post, while the tone of the shaded area represents the number of posts mentioning no tickers (the darker the shade, the higher the number).
Starting from January, most of the highest ranking posts are about GME, with a smaller fraction of posts mentioning other tickers and a progressively growing number of posts mentioning no tickers at all. 
Posts written by the three opinion leaders regularly appear amongst the top rated content, this outlines that once a user reaches such position, it will likely be maintained in time. 
To further inspect how the behavior of users changed after the short squeeze, we display in the middle panel of Figure \ref{fig4}C the number of posts written in the entire time spanned by our data and the average number of comments written per user, which can be regarded as a proxy for the interest of users in the content posted in the community. We see a steady decline, denoting the reduction of engaging conversations, even if the number of posts created in the post-GME short squeeze period is higher. 
Finally, the bottom panel of Figure \ref{fig4}C presents the monthly averages of quantities describing how users interact: the number of direct replies (\emph{first children}) of a comment  diminishes, indicating once again the lack of lengthy discussions; and so does the number of jargon terms used in writing posts, which reflects a user's affinity with the community \cite{wsb_positions_ban,lucchini_committed}. On the other hand the change in jargon concerning comments is not as appreciable, given its already low value due to the shorter length of the text. For a discussion on how these changes affected the topological structure of the network, see section \ref{assortativity} of the Supplementary materials.
\newpage

\section*{Conclusions}
Understanding human behavior is a difficult challenge, but data retrieved from social media may come to our aid. The information gathered through the analysis of user roles and their evolution plays an important part in deepening our understanding of the subject. As this concept is vast and undefined, studying how individuals act on different social networks helps us in finding behavioral patterns that can be generalized to a broader environment. One of the aims of this work is to help fill in the gap towards this final goal, by examining the Reddit community of WallStreetBets and its time evolution, observing how it responds to engaging events and increasing popularity. 

Reddit provides the perfect environment for these types of analyses, as on the one hand the presence of anonymity facilitates a looser behavior of users, on the other hand a better identification of opinion leaders is possible, since it overcomes topic limitation and prejudices that might be present when the identity of a user is known.

In this work, we unveiled the patterns of online behavior for average users through a semi-supervised procedure, and portrayed their time evolution. 
Furthermore, by looking at users whose behavior deviates considerably from average, we could recognise different social roles -- including opinion leaders.
Three main opinion leaders are identified, each linked to a specific stock investment. We analyze their behavior and find shared patterns that contribute to their popularity on WSB: persistent posting behavior, extreme investments (YOLOs) and successful gains. Although when considering the mass-coordination event that took place, only one of these users is truly relevant and was able to shift the focus of the community to his chosen stock: GME.

After the GME short squeeze, WSB experienced a massive inflow of subscribers. In this period, we found no actual investment-related opinion leader. Moreover, by looking at the contents produced by the community, we can witness a drastic change whereby investment-oriented discussions were outnumbered by meme shares and shorter, less focused conversations.

Our framework of analysis may be applied to other online social networks as well as off-line social domains, such as a friendship network, by adapting the feature vectors with the users' traits that are most relevant to each situation.
Nonetheless, in the particular case of WallStreetBets users, this work sheds light on social dynamics in an online environment with anonymity, identifying user roles and investigating for the first time the factors that contribute to becoming opinion leaders on Reddit, describing the features needed for users to obtain such a role, paving the way for similar analyses to be carried out in other settings.

\section*{Methods}

\subsection*{Dataset}
The dataset was collected from Pushshift \cite{pushshift}, an API that regularly copies and saves data from Reddit and other social networks. We used Pushshift's API to query all comments and posts from the subreddit WallStreetBets from August 1, 2020 to July 31, 2021. The dataset is then divided in two periods: before the GME short squeeze (Aug 2020 - Jan 2021) and after the short squeeze (Feb 2021 - Jul 2021). We treat each dataset separately but the same procedures are applied to both.

First, all comments and posts from known Reddit bots were removed.  Then, to select only active users during the considered timespan, we applied two separate filters. We consider the set of users who wrote more than 2 posts in the overall time period, the set of users who wrote more than 20 comments, and take their union to obtain a total of 103 597 unique users for the first period, and 112 661 for the second period. The final dataset comprises only posts and comments by these users, so it contains 12 272 782 comments and 603 106 posts for the first period, and 9 809 208 comments and 224 469 posts for the second period. 
Data regarding screenshots of the actual investment of a user were retrieved from the posts' metadata or by manual inspection of the specific post on WSB. 
The data relative to stock prices was retrieved from Yahoo Finance (\url{finance.yahoo.com}).

\subsection*{User networks and features}\label{methods-features}
The data is divided into weekly subsets by applying a 7-day moving window and shifting it by one day each time, so as to get a 6-day overlap between subsequent subsets. For simplicity, each subset will be referred to as a \emph{week}. For each user the following 16 features were computed over the given week, so in each week a single user is represented by a 16-dimensional array containing the (standardized) feature values.

\begin{table}[h!]
	\centering
    \begin{tabular}{ll}
    \toprule
    {\textbf{Comment Feature}} & {\textbf{Description}}\\
    \midrule
       \emph{number of comments} & number of comments made by the user\\
       \emph{score of comments} & average score of comments made by the user\\
       \emph{entropy of comments} & average value of the text complexity of the comments made by the user\\
       \emph{jargon of comments} & average number of jargon terms used in comments\\
       \emph{sentiment of comments} & average value of the sentiment of comments made by the user\\
       \emph{first children} & average number of comments that are direct replies to a comment by the user\\
    \bottomrule
    \end{tabular}
\label{tab_comm_feat}
\end{table}
\newpage
\begin{table}[h!]
	\centering
    \begin{tabular}{ll}
    \toprule
    {\textbf{Post Feature}} & {\textbf{Description}}\\
    \midrule
       \emph{number of posts} & number of posts made by the user\\
       \emph{score of posts} & average score of posts made by the user\\
       \emph{entropy of posts} & average value of the text complexity of the posts made by the user\\
       \emph{jargon of posts} & average number of jargon terms used in posts\\
       \emph{sentiment of posts} & average value of the sentiment of posts made by the user\\
       \emph{post comments} & average number of comments under a given post made by the user\\ 
    \bottomrule
    \end{tabular}
\label{tab_post_feat}
\end{table}

\begin{table}[h!]
	\centering
    \begin{tabular}{ll}
    \toprule
    {\textbf{Network Feature}} & {\textbf{Description}}\\
    \midrule
       $k_{out}$ & out-degree of the user in the weekly network\\
       $k_{in}$ & in-degree of the user in the weekly network\\
       $s_{out}$ & out-strength of the user in the weekly network\\
       $s_{in}$ & in-strength of the user in the weekly network\\
    \bottomrule
    \end{tabular}
\label{tab_net_feat}
\end{table}

The score of a post or comment is defined as the number of "upvotes" minus the number of "downvotes" it receives from users.\\ 
The entropy of the posts/comments by user $\alpha$ is computed as
\begin{equation}
    E_{\alpha} = -\sum_{i = 1}^{N} p_{i}^{\alpha} \cdot ln(p_{i}^{\alpha}) / ln(N^{\alpha})
    \label{eq.ent}
\end{equation}
where $p_{i}^{\alpha}$ is the number of times user $\alpha$ uses word $i$ divided by the total number of words used by $\alpha$ in the given week, $N^{\alpha}$. We use as normalization the maximum possible value of entropy, $log(N^{\alpha})$, when the user uses every word with equal probability $1/N^{\alpha}$, to make eq. \eqref{eq.ent} independent from the number of comments and posts written by the user. Entropy is set to 0 when the user employs only 1 word or when the same word is used multiple times.

The root words of jargon terms selected to compute the corresponding feature are: \emph{retard, autist, ape, degenerate, moon, bull, bear, bagholder, btfd, diamond, paper, dd, stonk, tendies, yolo, hodl} \cite{wsb_jargon}. These words are commonly used in the community's language, far before the GME short squeeze. Thus we can say that these jargon terms have not evolved in time, as we have already shown in our previous work\cite{sel-induced_reddit}.

Sentiment of comments and posts is evaluated using VADER (Valence Aware Dictionary and sEntiment Reasoner) \cite{Hutto_Gilbert_2014}, an algorithm that assigns a piece of text with a compound score between $-1$ (very negative) and $+1$ (very positive). VADER is sensitive to both the polarity and intensity of the text, taking into account punctuation and word shape (ALL CAPS) used to add emphasis, degree modifiers that alter intensity (boosters such as \emph{very} and dampeners such as \emph{kind of }), slang and acronyms. VADER is based on a lexicon of words and emojis, each with an associated score ranging from $-4$ to $+4$ according to its meaning (from negative to positive). As done in our previous study, we adapt VADER to the typical jargon and sarcasm of WSB users by adding to its lexicon particular words (see \cite{sel-induced_reddit} for the detailed procedure).

The weekly networks from which the network features are computed are constructed by considering the reply-to interactions, a method commonly used when dealing with online social networks \cite{reply-to_networks,networks_slashdot}. Given two users A and B, if A has replied to a comment or post written by B, then a directed link from A to B is added. The weight of the link represents how many times A has replied to something written by B. So the resulting networks are directed, weighted and with no self-links.

\subsection*{DBSCAN and spectral clusteirng} \label{Dbscan_methods}
Since some of our 16 features may be correlated, we apply PCA to the matrix of users' feature vectors. This allows us to focus only on the top 10 PCA components -- which on average keep around 95\% of the variance of the original features (see Supplementary section \ref{supp_PCA} for further details).

Hence both DBSCAN and spectral clusteirng are applied to these 10-component PCA projection of the user's feature vectors, meaning that a user is represented as a point in this 10-dimensional space. 
DBSCAN (Density-Based Spatial Clustering of Applications with Noise) \cite{dbscan} is a clustering algorithm that uses density of data points in a given region to detect clusters. It groups into same clusters points that are packed together, labeling as noise those that lie in low-density regions. It works by defining two parameters, \emph{MinPts} and $\epsilon$, that are used to classify points in three categories: \emph{core points}, if at least \emph{MinPts} points are within a distance $\epsilon$; \emph{reachable points}, if they lie within distance $\epsilon$ from a core point; \emph{noise points}, all those that are not reachable from any point. 
Core points and reachable points are the ones that form clusters. We use \emph{MinPts} = 15 and $\epsilon$ = 1 (see Supplementary Materials \ref{dbscan_paramerters} for a robustness analysis).

We also used spectral clustering \cite{spectral_clustering} to further validate the clustering obtained with DBSCAN. First, the pairwise distances between all users were computed and stored in a matrix $D$, then the affinity matrix was computed as $S = \exp(-D^2)$. All matrix entries whose value is below 0.5 are set to 0, and a weighted graph is built using the remaining matrix elements. The connected components of the graph are the user clusters, while all isolated nodes along with all the connected components with less than 5 elements are set to noise.

Both methods were implemented using the Python package Scikit-learn \cite{scikit-learn} and give very similar results (see Supplementary Material \ref{supp_ARI}).
Note that to make the analysis robust against the growing size of the community, we performed the clustering on 100 different samples of 10 000 users, and consider the mean and standard deviation of all quantities obtained.

\subsection*{Outliers with iterative DBSCAN clustering} \label{outliers}
In addition to being a popular clustering algorithm, DBSCAN has also been used for anomaly detection \cite{anomaly_temperature,anomaly_terminal,anomaly_bitcoin}. In fact, by focusing on points that are classified as noise, we can zoom-in on those individuals whose behavior deviates from that of the average user. To select only the most persistent outliers we follow a two-step procedure. 
1) For each week we iteratively apply DBSCAN 12 times, each time we exclude all points who were assigned to clusters and only consider those labelled as noise, to which we apply again the clustering algorithm. At each iteration the parameter \emph{MinPts} is fixed to 10, while $\epsilon$ varies from 1 to 12, adapting to the decreasing density of the dataset. After this first step we are left with only the furthest outliers for each week. 
2) From these outliers we select those who appear for more than 15 weeks (out of the 177 total) for the first half of the dataset, and for more than 20 weeks (out of the 175 total) for users in the second half of the dataset (since we have more users we need a stronger filter). In this way we select those users who show an outlier behavior persistently in time.

\subsection*{Ticker entropy} \label{ticker-entropy}
Since WallStreetBets is a community of traders, where most of the conversations revolve around stocks, we define the \emph{ticker entropy} as a measure of the heterogeneity of stocks discussed in the community. We consider the text of posts, as they contain much of the the information about the preferences of the community, and select all the uppercase regular expressions of 2 to 5 characters. From all the expressions found we manually validated those that actually corresponded to ticker symbols, and excluded all those with ambiguous meaning (FROG, YOU, etc...). In the end we are left with $M = 157$ stock tickers that have been discussed in the time span considered. For each week we compute the probability $p_{k}$ of ticker $k$ by dividing the total number of mentions of $k$ by the total number of ticker mentions in the given week. We then compute the Shannon entropy of the discrete ticker distribution of week $tau$ as
\begin{equation}
    H_{\tau} = -\sum_{k = 1}^{M}p_{k}^{\tau}\cdot ln(p_{k}^{\tau})
\end{equation}

\section*{Author contributions statement}
A.D., A.M. and G.P. gathered the data. A.M. performed the analysis and realised the figures. A.M., R.D.C. and G.C. designed the analysis. All the authors discussed the results, wrote the paper and approved the final manuscript.

\section*{Additional information}
\textbf{Competing interests} The authors declare no competing interests. 

\section*{Data availability}
Raw Reddit data are available from Pushshift \cite{pushshift} (subscription required). 
The data and code used for the analyses and the figures of the manuscript will be available on the GitHub \url{https://github.com/ComplexConnectionsLab/rise_and_fall_of_wallstreetbets} after acceptance of the manuscript.

\section*{Acknowledgments}
R.D.C. acknowledges Sony Computer Science Laboratories Paris, for hosting him during part of the research.
G.C. acknowledges support from ``Deep `N Rec'' Progetto di Ricerca di Ateneo of University of Rome Tor Vergata.

\bibliography{biblio}

\newpage

\section*{\LARGE The rise and fall of WallStreetBets: social roles and opinion leaders across the GameStop saga}
\author{\textbf{Anna Mancini, Antonio Desiderio, Giovanni Palermo, Riccardo Di Clemente, Giulio Cimini}}

\section*{\LARGE Supplementary Materials}

\setcounter{figure}{0}
\setcounter{table}{0}
\setcounter{page}{1}
\makeatletter
\renewcommand{\thesection}{S\arabic{section}}
\renewcommand{\thetable}{S\arabic{table}}
\renewcommand{\thefigure}{S\arabic{figure}}

\bigskip

\section{List of moderators and bots}\label{sm_bots}
The current list of moderators is always available on the page of the subreddit, but many of the users who were moderators during the period we considered are not anymore. In order to identify the active moderators during the six months spanned by the first period of our data, we checked all comments and posts looking for users who referred to themselves as moderators or were referred as such by other members. Crosschecking the users we found with their activity on WSB, all those who banned users from the subreddit during those 6 months are considered to be moderators, since this is a task that only a moderator can perform.
Concerning bots, we found them by looking for all users whose username ends with "bot" and manually validating each result by looking at the comments they wrote.

\begin{table}[h!]
	\centering
    \begin{tabular}{l}
    \toprule
    {\textbf{Username of moderators}}\\
    \midrule
       Only1parkjisung\\
       wallstreetboyfriend\\
       GoBeaversOSU\\
       OPINION\_IS\_UNPOPULAR\\
       zjz\\
       ItradeBaconFutures\\
       grebfar\\
       jartek\\
       stormwillpass\\
       premier\_\\
       CHAINSAW\_VASECTOMY\\
       Dan\_inKuwait\\
       sdevil713\\
       CallsOnAlcoholism\\
       theycallmeryan\\
    \bottomrule
    \end{tabular}
    \hspace{2cm}
    \begin{tabular}{l}
    \toprule
    {\textbf{Username of bots}}\\
    \midrule
    AutoModerator\\
    WhoDidThisBot\\
    DuplicatesBot\\
    anti-gif-bot\\
    alternate-source-bot\\
    gifv-bot\\
    hug-bot\\
    AnimalFactsBot\\
    \_youtubot\_\\
    tweettranscriberbot\\
    HelperBot\_\\
    CommonMisspellingBot\\
    tupac\_cares\_bot\\
    RedditSilverRobot\\
    WSBVoteBot\\
    RemindMeBot\\
    Generic Reddit Bot\\
    ReverseCaptioningBot\\
    LimbRetrieval-Bot\\
    NoGoogleAMPBot\\
    RepostSleuthBot\\
    GetVideoBot\\
    CouldWouldShouldBot\\
    \bottomrule
    \end{tabular}
    \caption{List of users who were moderators during the six months spanned by our data (left column) and list of bots removed from our dataset (right column).} 
\label{tab_mods}
\end{table}

\section{Feature distributions}
We present a joyplot of the weekly distributions of all features we have considered.
\newpage
\begin{figure}[h!]
    \centering
    \includegraphics[width=\textwidth]{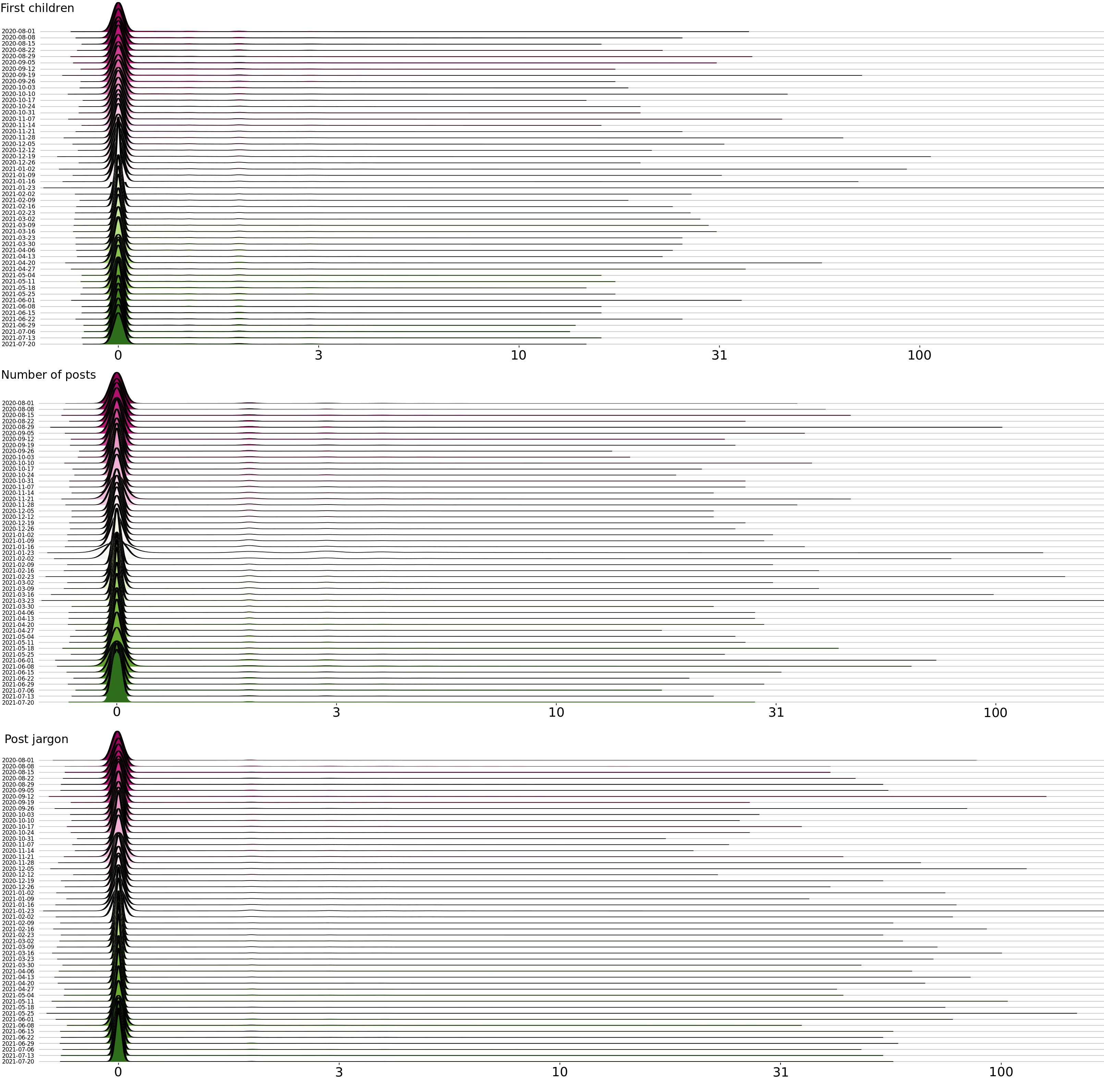}
\end{figure}
\newpage
\begin{figure}[h!]
    \centering
    \includegraphics[width=\textwidth]{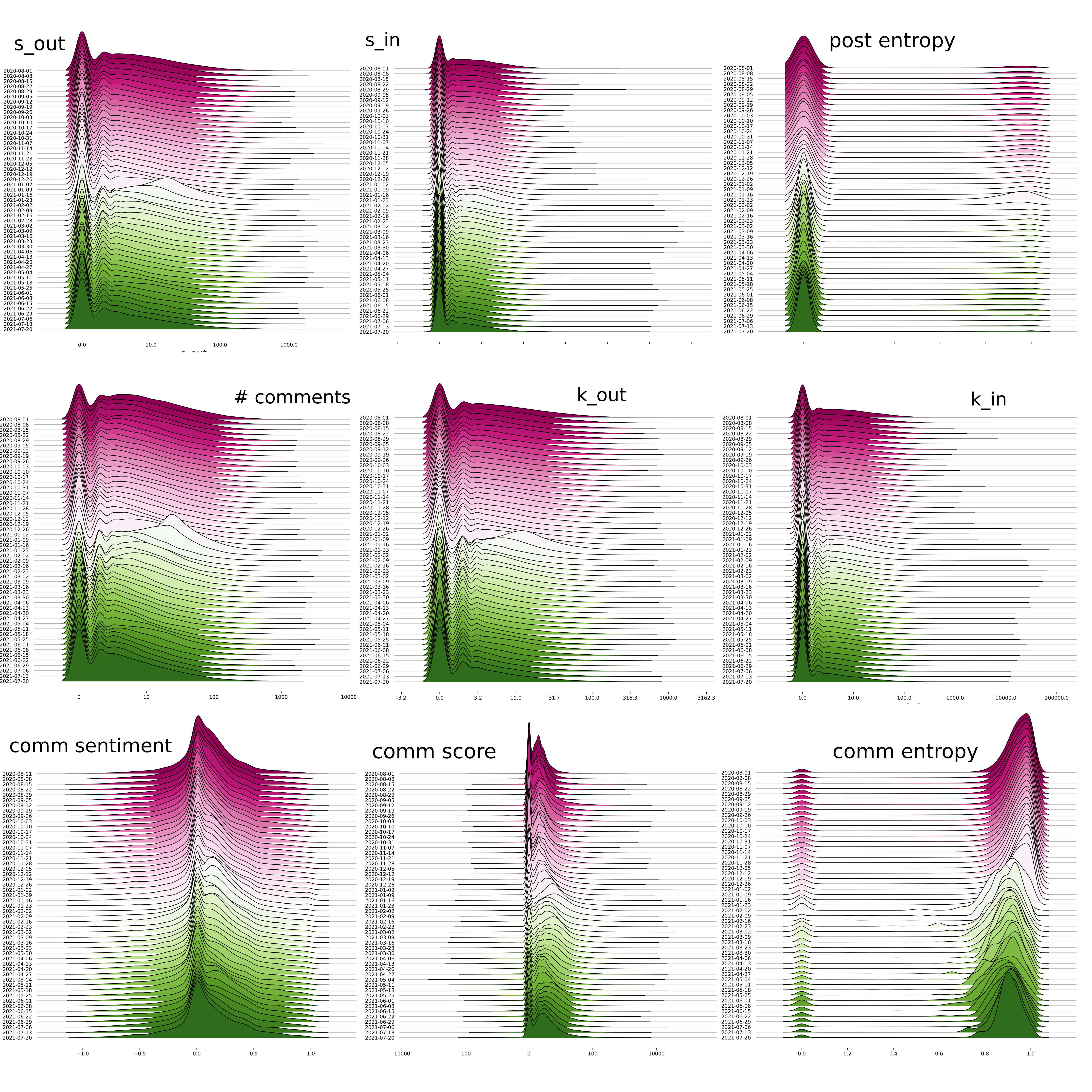}
\end{figure}
\newpage
\begin{figure}[h!]
    \centering
    \includegraphics[width=\textwidth]{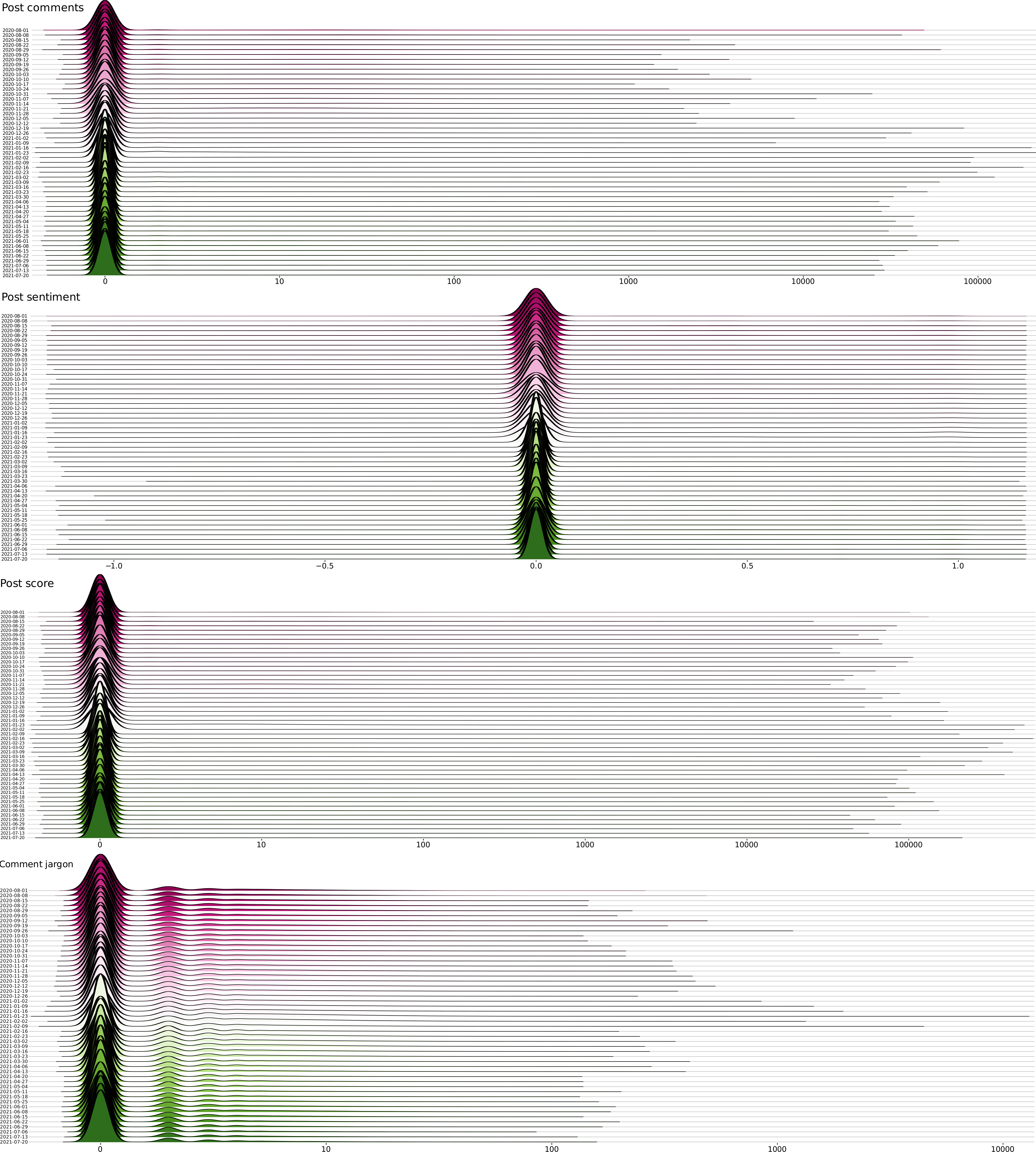}
\end{figure}
\newpage

\section{Moderator behavior} \label{mod_beh}
\begin{figure}[ht!]
    \centering
    \includegraphics[width=0.8\textwidth]{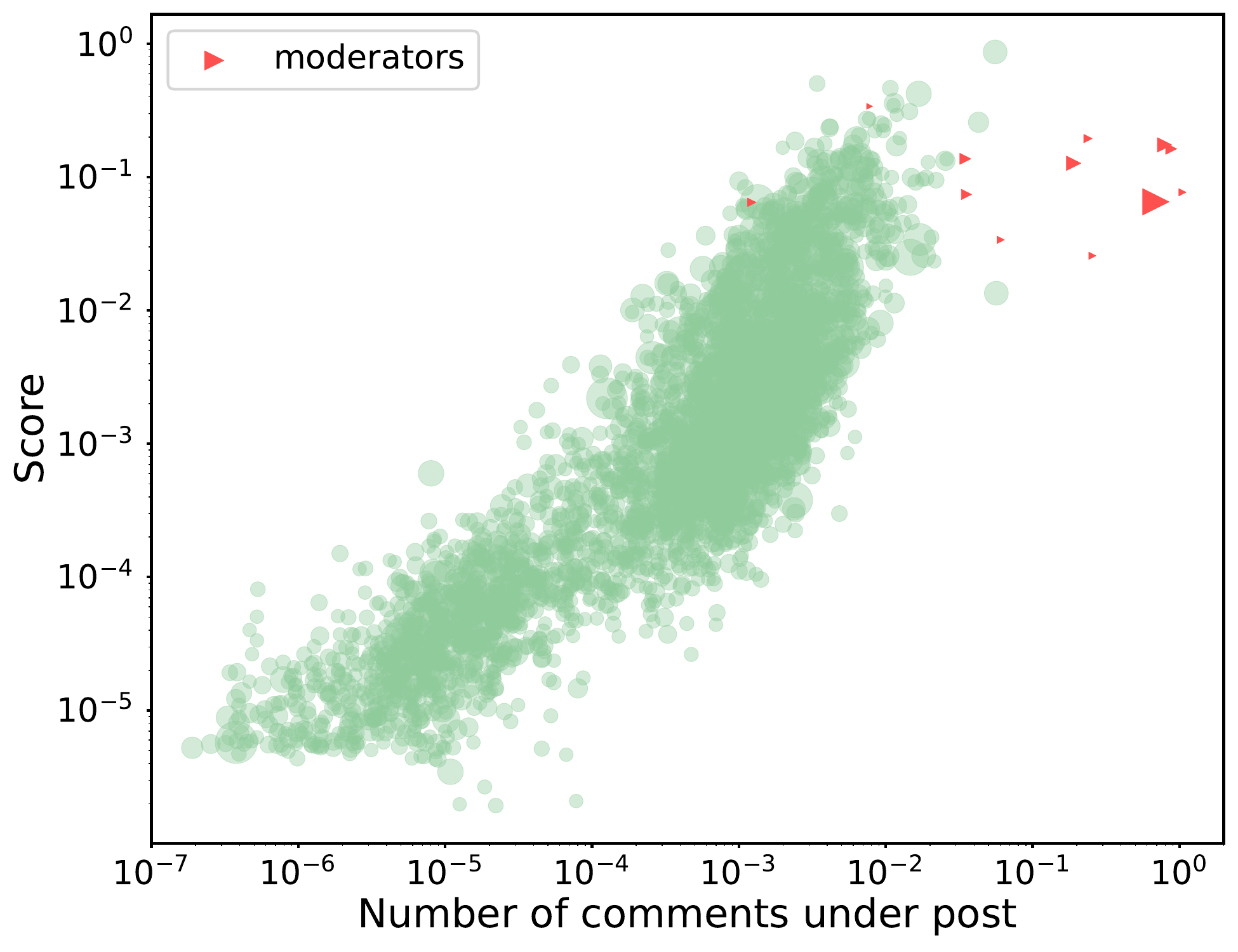}
    \caption{\textbf{Moderator behavior} In green are all users with more than 10 posts, the pink triangles are the moderators of the community. We see that moderators do not follow the trend of the "average" user, but instead lie in upper right section of the plot, due to the fact that they are the ones that write the \emph{Daily Thread} and other highly commented threads.}
    \label{fig_mods}
\end{figure}

Concentrating on the first part of our dataset, for which we have identified most of the moderators, by analyzing each one's commenting and posting behavior we find that most of them stick to their role, interacting on the community only to ensure that each member respects the rules and never talking about their personal investments. They are the ones that make the \emph{Daily Thread}, a post where users are encouraged to write by the rules of WSB. Because of this, Figure \ref{fig_mods} shows how moderators have a completely different trend when compared to all other users in the community.\\

\section{User filter}
\begin{figure}[ht!]
    \centering
    \includegraphics[width=\textwidth]{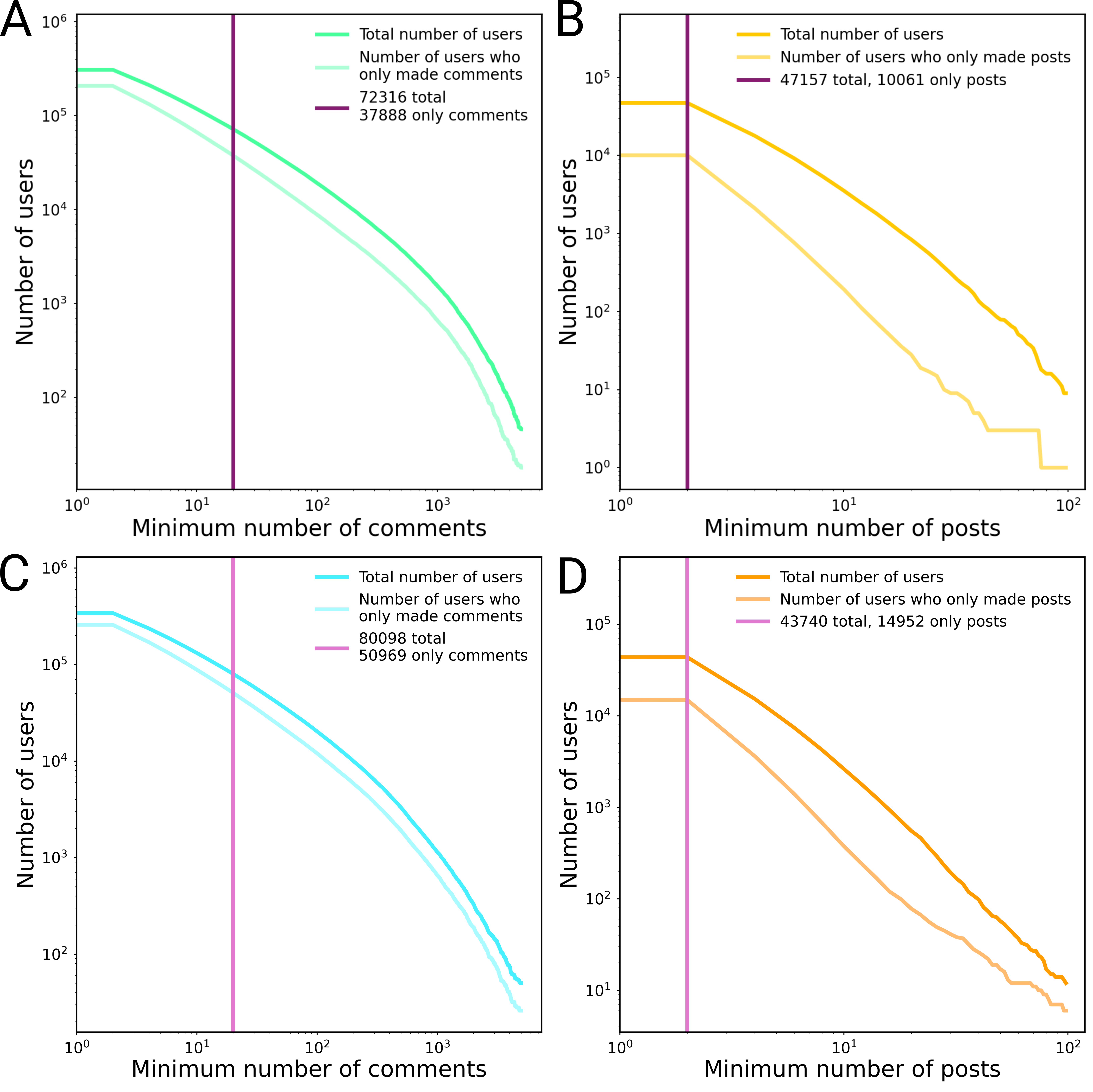}
    \caption{\textbf{User filter} A) Filter on number of comments for the period August 2020 - January 2021: number of users as a function of the minimum number of comments written, both for the total number of users and for users who only wrote comments. Threshold chosen at a minimum of 2 comments per user. B) Filter on number of posts for the period August 2020 - January 2021: number of users as a function of the minimum number of posts written, both for the total number of users and for users who only wrote posts. Threshold chosen at a minimum of 2 posts per user. The users resulting from the two filters are merged and duplicate users discarded, resulting in 103 597 distinct users. C) Filter on number of comments for the period February 2021 - July 2021: number of users as a function of the minimum number of comments written, both for the total number of users and for users who only wrote comments. Threshold chosen at a minimum of 2 comments per user. D) Filter on number of posts for the period February 2021 - July 2021: number of users as a function of the minimum number of posts written, both for the total number of users and for users who only wrote posts. Threshold chosen at a minimum of 2 posts per user. The users resulting from the two filters are merged and duplicate users discarded, resulting in 112 663 distinct users.}
    \label{fig_userfilter}
\end{figure}

Given the large number of users who have interacted on WSB in both periods we considered (pre- and post-short squeeze), before starting the analyses we decided to apply a filter to eliminate all those who were less active. We do so by implementing two separate filters, one on comments and one on posts, as is displayed in Figure \ref{fig_userfilter} A and B for the first part of the dataset, and C and D for the second one, where the purple vertical line indicates where the filter is set. The first group is made up of all users who have written more than 20 comments, the second of all those who have written more than 2 posts, the two groups are then merged and duplicate elements discarded to obtain the two final datasets of distinct users: 103 597 users for the first period and 112 663 for the second period.
\newpage
\newpage
\section{Active users vs subscribers}\label{1_rule}
The so-called 1\% rule has been seen across different social networks, and Reddit is no exception. What this rule states is that only 1\% of the users are the ones who actually contribute to conversations and create new information, the remaining 99\% are lurkers, people who read but do not interact. What is somewhat surprising is that such rule still stands even when the community starts to grow due to the success of the short squeeze. One would imagine that at least in this case most new subscribers must also be eager to engage in conversations, given the popularity WSB had earned. As Figure \ref{1_rule} shows this is not true, since the fraction of active users (dashed purple line) is always around 0.01, and peaks at 0.03 at the beginning of January only to drop down again to an even lower level.

\begin{figure}[ht!]
    \centering
    \includegraphics[width=0.9\textwidth]{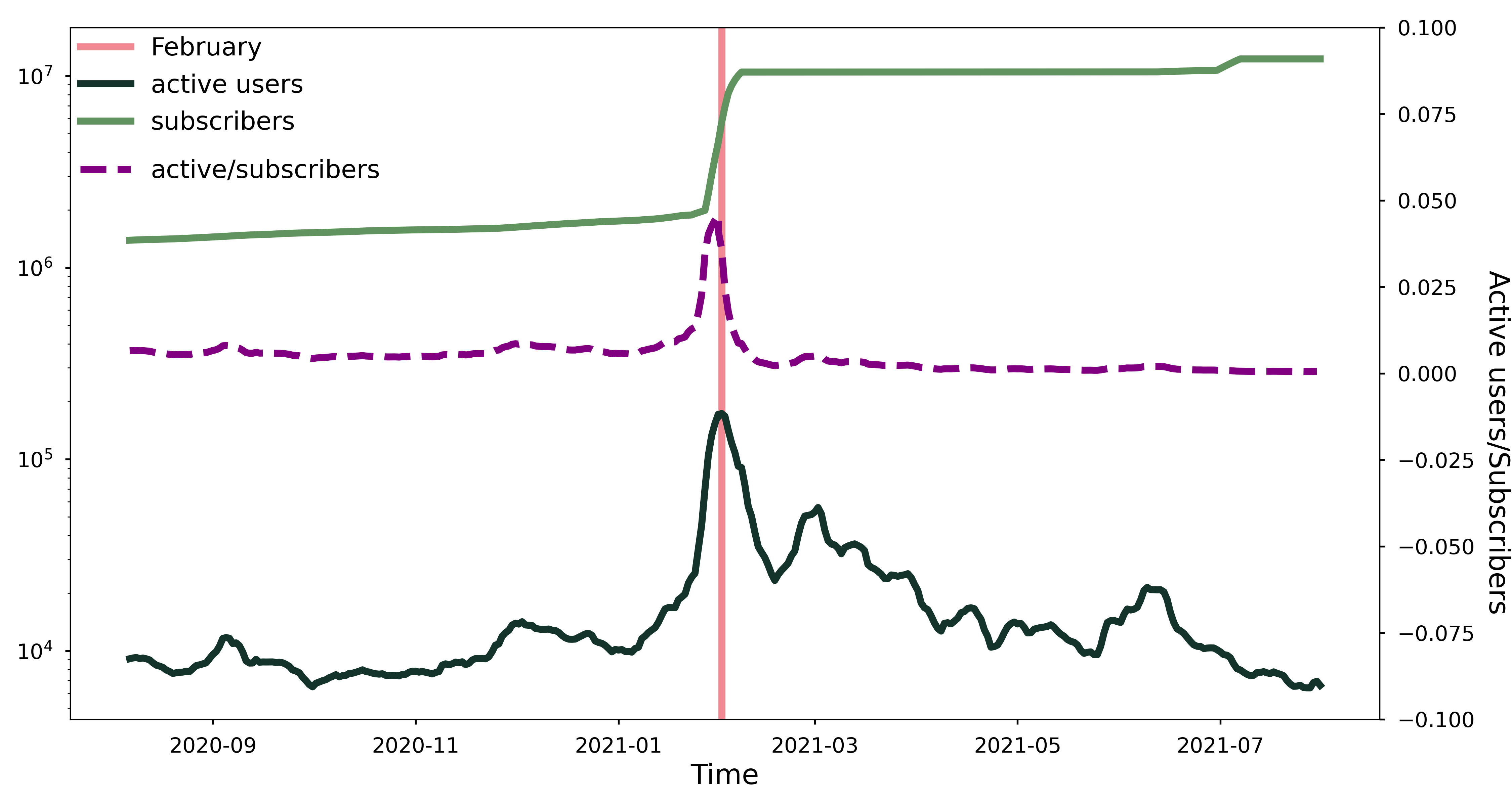}
    \caption{\textbf{1\% rule} The solid lines represent the number of subscribers on WSB in time (green) and the number of daily active users (blue), were a 7-day moving average was applied to the latter in order to remove the circadian oscillations. The dashed purple line is the fraction of active users on daily subscribers, and it oscillates around a value of 0.01, thus confirming also in this scenario the 1\% rule.}
    \label{fig_1_percent}
\end{figure}

\section{PCA explained variance and feature importance} \label{supp_PCA}
\begin{figure}[ht!]
    \centering
    \includegraphics[width=\textwidth]{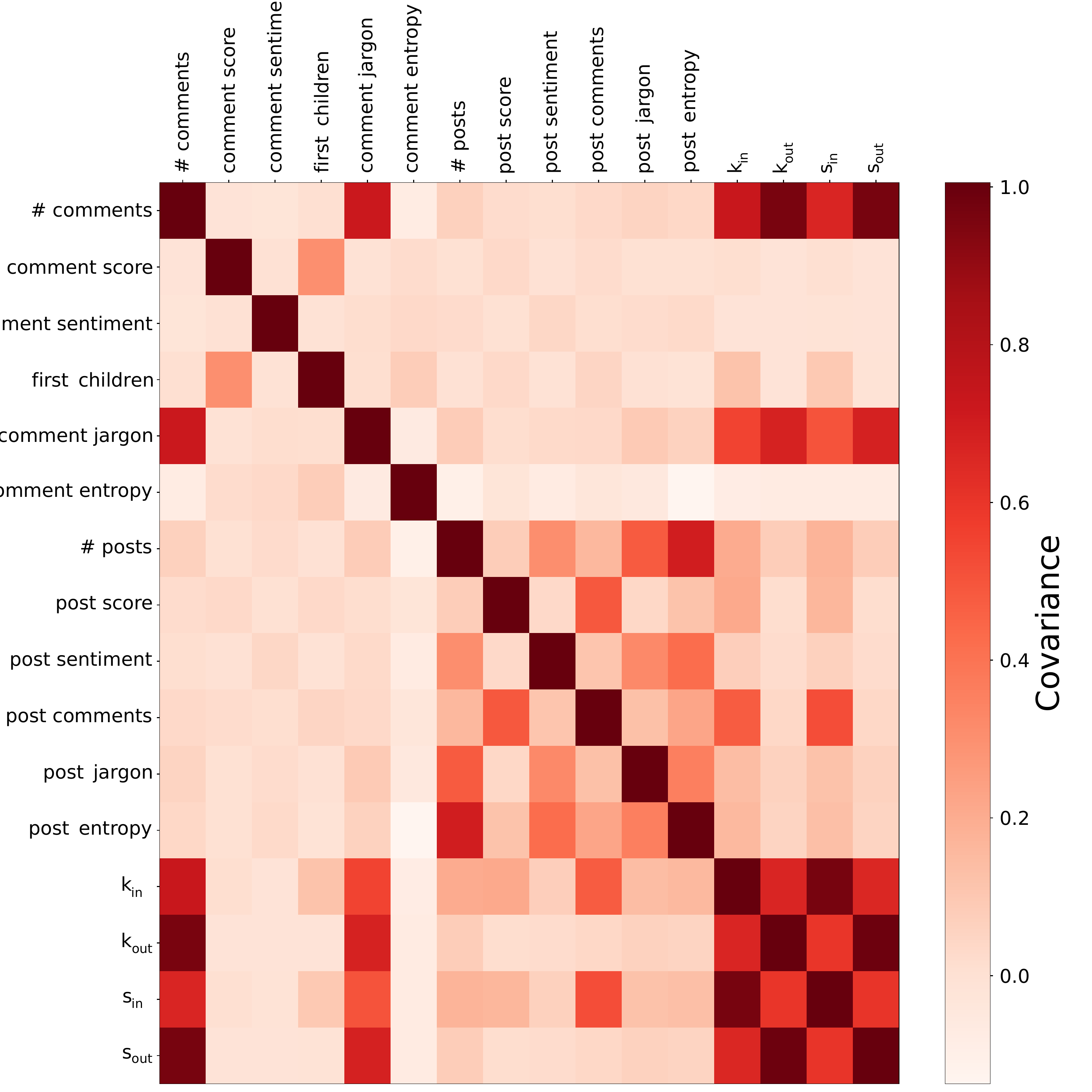}
    \caption{\textbf{Average variance-covariance matrix} Variance-covariance matrix computed as the average over all 177 weeks. The matrix shows how some groups of features are strongly correlated, as we see by the darker shade of red in some of the cells.}
    \label{fig_correlation}
\end{figure}

\begin{figure}[ht!]
    \centering
    \includegraphics[width=\textwidth]{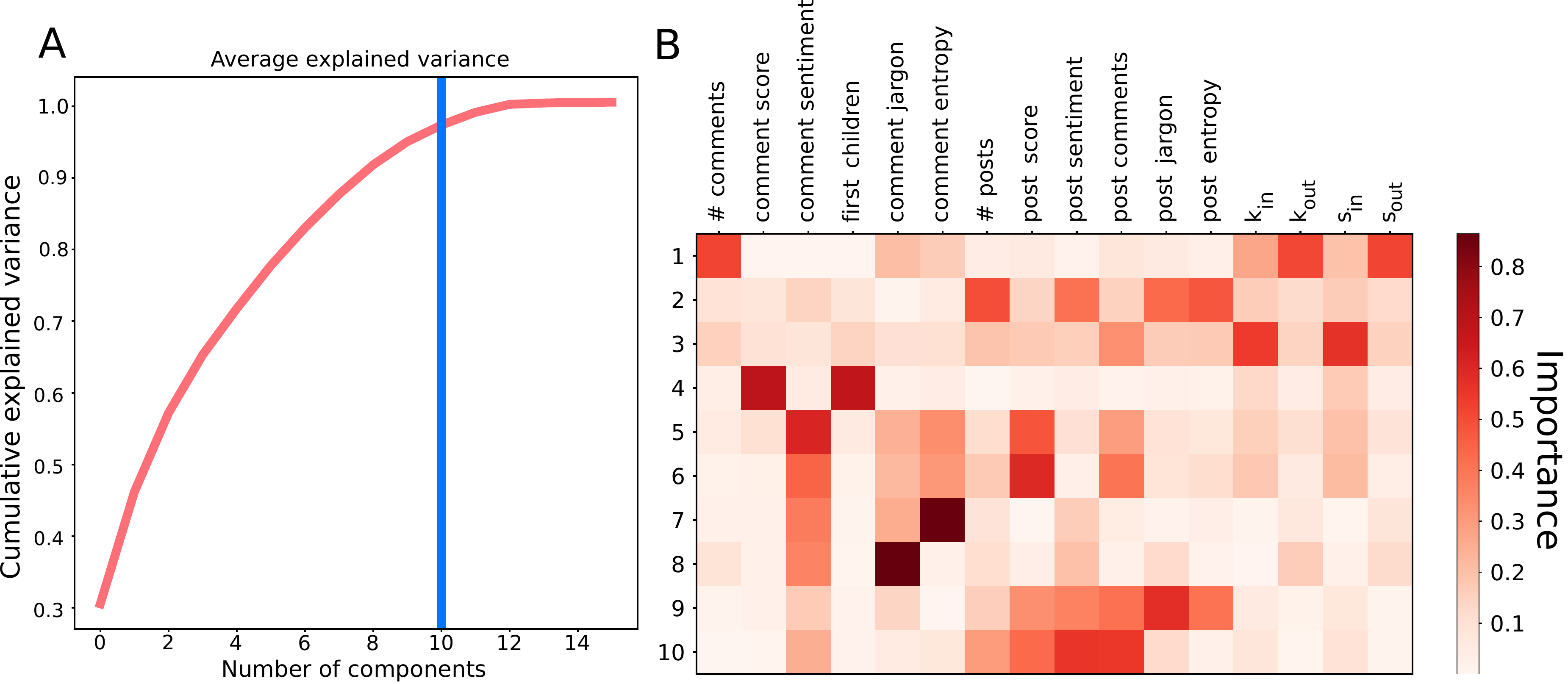}
    \caption{\textbf{PCA components and feature importance} A) Cumulative sum of the average explained variance over all 177 weeks. The vertical line indicates our chosen number of components to keep, 10, in order to have on average 95\% of the explained variance. B) Average feature importance over all 177 weeks for each principal component. The darker the shade of red, the more important the feature in the given component.}
    \label{fig_pca}
\end{figure}

Figure \ref{fig_correlation} shows the variance-covariance matrix of all features as an average over all 177 weeks (first period only is considered), and it allows to grasp the intrinsic correlation between some of the features we have considered. To remove this effect and perform our analyses in a lower dimensional space, we apply PCA to all feature vectors of all weeks and work directly on this projection. Figure \ref{fig_pca} A shows the explained variance for each principal component considered, we chose to keep 10 components in order to have around 95\% of explained variance. Figure \ref{fig_pca} B shows how the feature importance is distributed in each principal component.
\newpage
\newpage

\section{DBSCAN parameter selection}\label{dbscan_paramerters}
\begin{figure}[hbt!]
    \centering
    \includegraphics[width=0.9\textwidth]{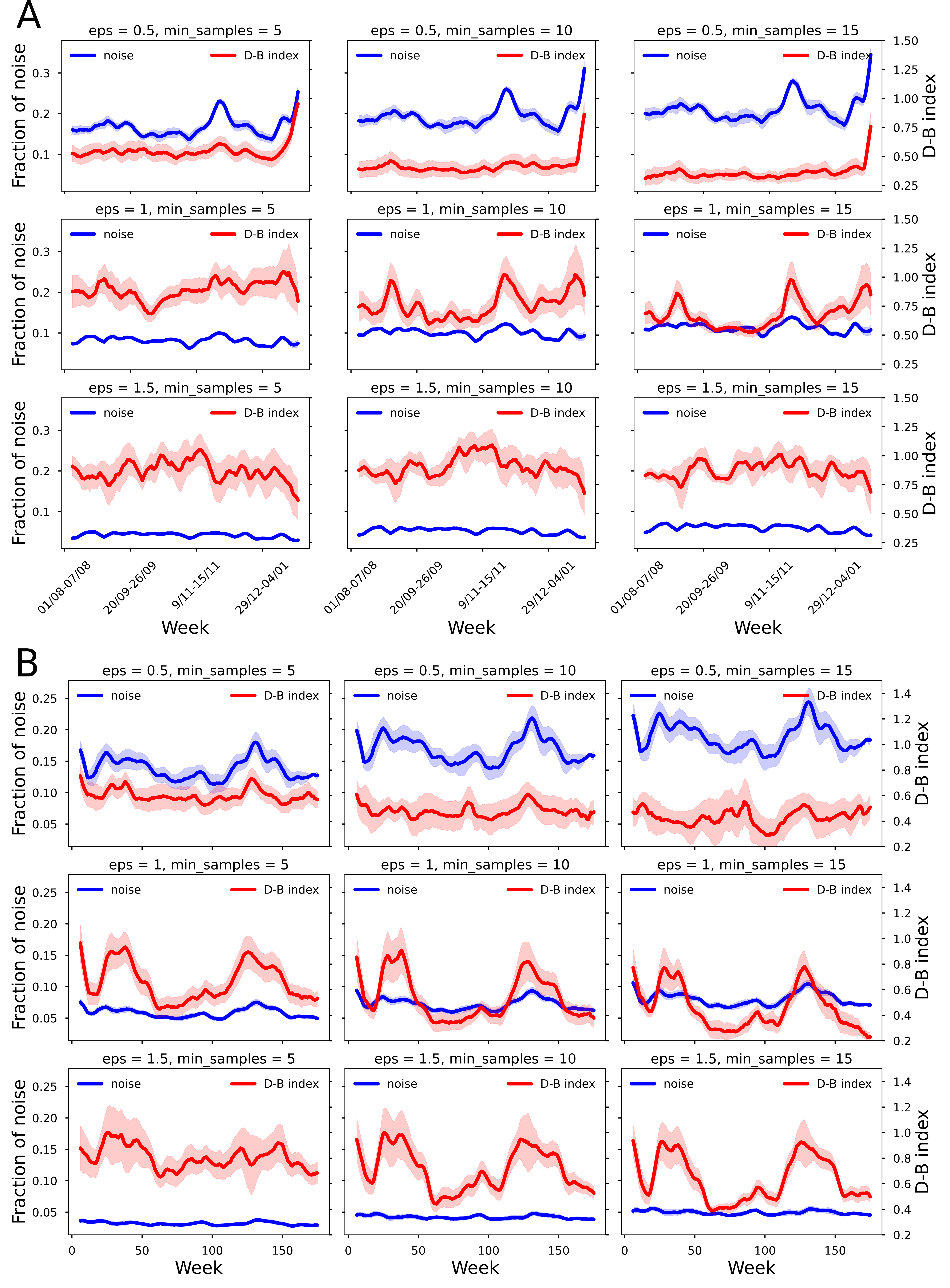}
    \caption{\textbf{Parameter evaluation} A) DBSCAN for the first dataset from August 2020 to January 2021, for 9 different sets of parameters, in red the Davies-Bouldin index, in blue the fraction of noisepoints. We can see that the configuration with the best set of parameters is the one where $\epsilon$ = 1 and \emph{MinPts} = 15. 7-day moving window applied to all time series. B) DBSCAN for the second dataset from February 2020 to July 2021, for 9 different sets of parameters, in red the Davies-Bouldin index, in blue the fraction of noisepoints. We can see that the configuration with the best set of parameters is the one where $\epsilon$ = 1 and \emph{MinPts} = 15. 7-day moving window applied to all time series.}
    \label{fig_parameters_dbscan}
\end{figure}

DBSCAN is a clustering algorithm that does not need a predetermined value for the number of clusters, however it requires two parameters \emph{MinPts} and $\epsilon$, that need to be carefully set to obtain the best results. To choose the optimum pair we perform DBSCAN, with the following values of the parameters, for all weeks of both our datasets:
\begin{itemize}
    \item $\epsilon$ = 0.5, \emph{MinPts} = 5
    \item $\epsilon$ = 0.5, \emph{MinPts} = 10
    \item $\epsilon$ = 0.5, \emph{MinPts} = 15
    \item $\epsilon$ = 1, \emph{MinPts} = 5
    \item $\epsilon$ = 1, \emph{MinPts} = 10
    \item $\epsilon$ = 1, \emph{MinPts} = 15
    \item $\epsilon$ = 1.5, \emph{MinPts} = 5
    \item $\epsilon$ = 1.5, \emph{MinPts} = 10
    \item $\epsilon$ = 1.5, \emph{MinPts} = 15
\end{itemize}
For each pair of values we perform 100 clusterings, each time picking a different random sample of 10 000 users. We choose the best set of parameters by looking at the ones that minimize two quantities: the number of noise points and the Davies-Bouldin index \cite{d-b_index}. While the former simply means that the most desirable clustering is the one where the majority of points are part of a cluster (and are thus not labelled as noise), the latter reaches a minimum when the separation between clusters is maximal and the variation within clusters has a minimum, thus identifying the best clustering outcome. Figure \ref{fig_parameters_dbscan} shows the clustering results for each pair of parameters over all weeks, where the shaded region is the standard deviation over the 100 different realizations. We can see how the best clustering realizations, over all weeks, are the ones corresponding to $\epsilon$ = 1 and \emph{MinPts} = 15, as both the Davies-Bouldin index and noise have minimum values.
\clearpage
\clearpage
\section{Adjusted Rand index} \label{supp_ARI}
\begin{figure}[hbt!]
    \centering
    \includegraphics[width=0.8\textwidth]{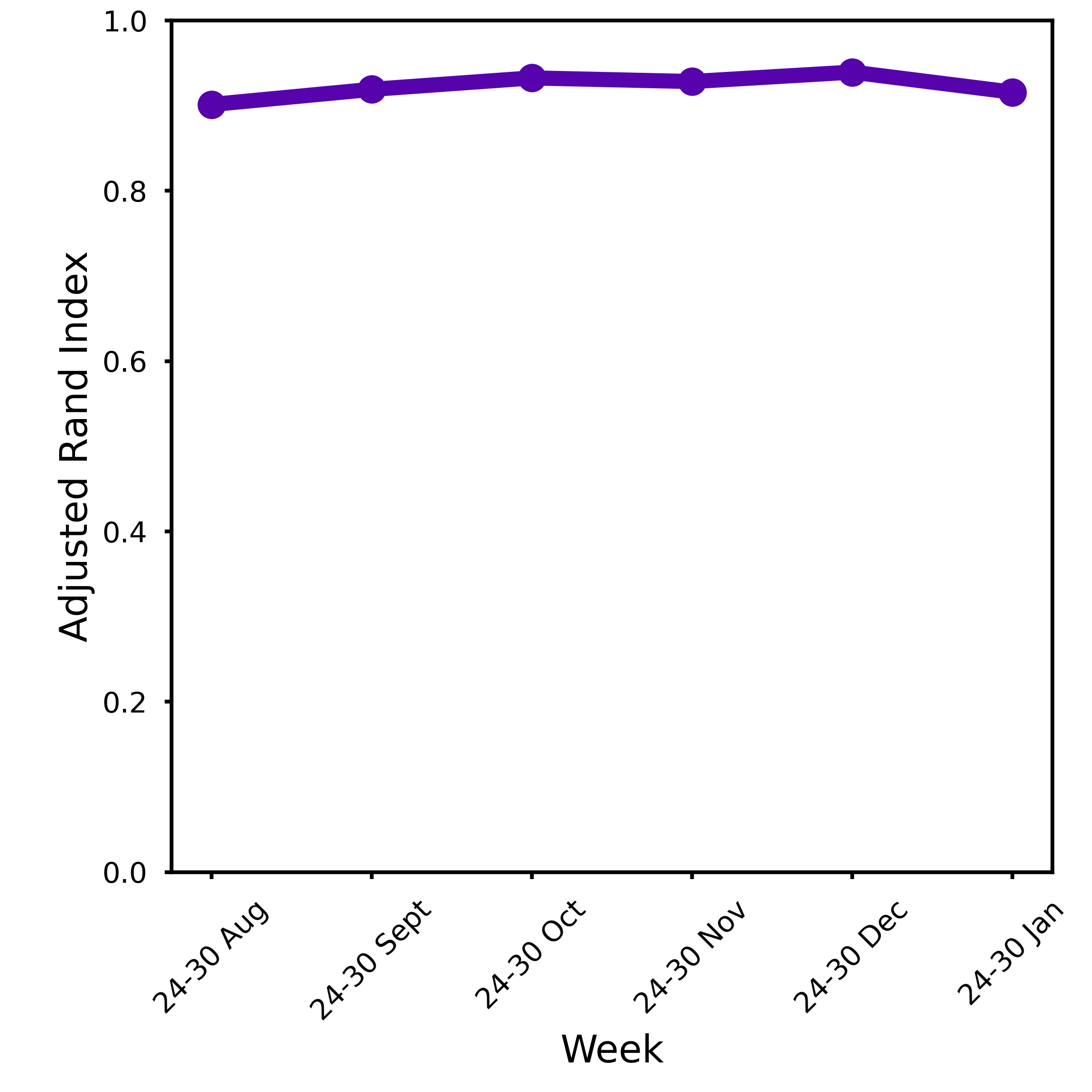}
    \caption{\textbf{Adjusted Rand index} Values of the Adjusted Rand index for 6 chosen weeks. The values are all very close to 1, indicating an almost perfect agreement between the DBSCAN clustering and the spectral clustering.}
    \label{fig_ari}
\end{figure}

The \emph{Rand index} is a measure of similarity between two data clusterings. Given a set of elements $S = \{s_1, s_2,..., s_n\}$ and two different clustering algorithms that partition $S$ into subsets $X = \{X_1, X_2,..., X_k\}$ and $Y = \{Y_1, Y_2,..., Y_l\}$, the Rand index is computed as
\begin{equation}
    RI = \frac{a + b}{\binom{n}{2}}
\end{equation}
where $a$ is the number of pairs of elements in $S$ that are in the same subset in $X$ and in the same subset in $Y$, $b$ is the number of pairs of elements in $S$ that are in different subsets in $X$ and in different subsets in $Y$ and the denominator is the total number of pairs. The rand index has a value between 0 and 1, where 0 indicates that the two clusterings do not agree on any pair of points and 1 indicating that they are exactly the same. However this measure does not take into account the fact that some agreement between the two clusterings can occur by chance. The \emph{adjusted rand index} takes this possibility into account and is computed as  
\begin{equation}
    ARI = \frac{RI - \mathbb{E}[RI]}{max(RI) - \mathbb{E}[RI]}
\end{equation}
where $\mathbb{E}[RI]$ is the expected value of the rand index between clusterings of a random model and $max(RI) = 1$. ARI values range between -1 and 1, were 1 indicates perfect agreement between the two clusterings,  0 indicates random agreement and -1 indicates that the two clusterings are completely different. Figure \ref{fig_ari} shows the value of the ARI for the two clusterings obtained with DBSCAN and spectral clustering on the first part of the dataset, and we see how for each week the two algorithms agree almost to perfection.
\newpage
\newpage
\section{Persistent users analysis}
\begin{figure}[hbt!]
    \centering
    \includegraphics[width=0.9\textwidth]{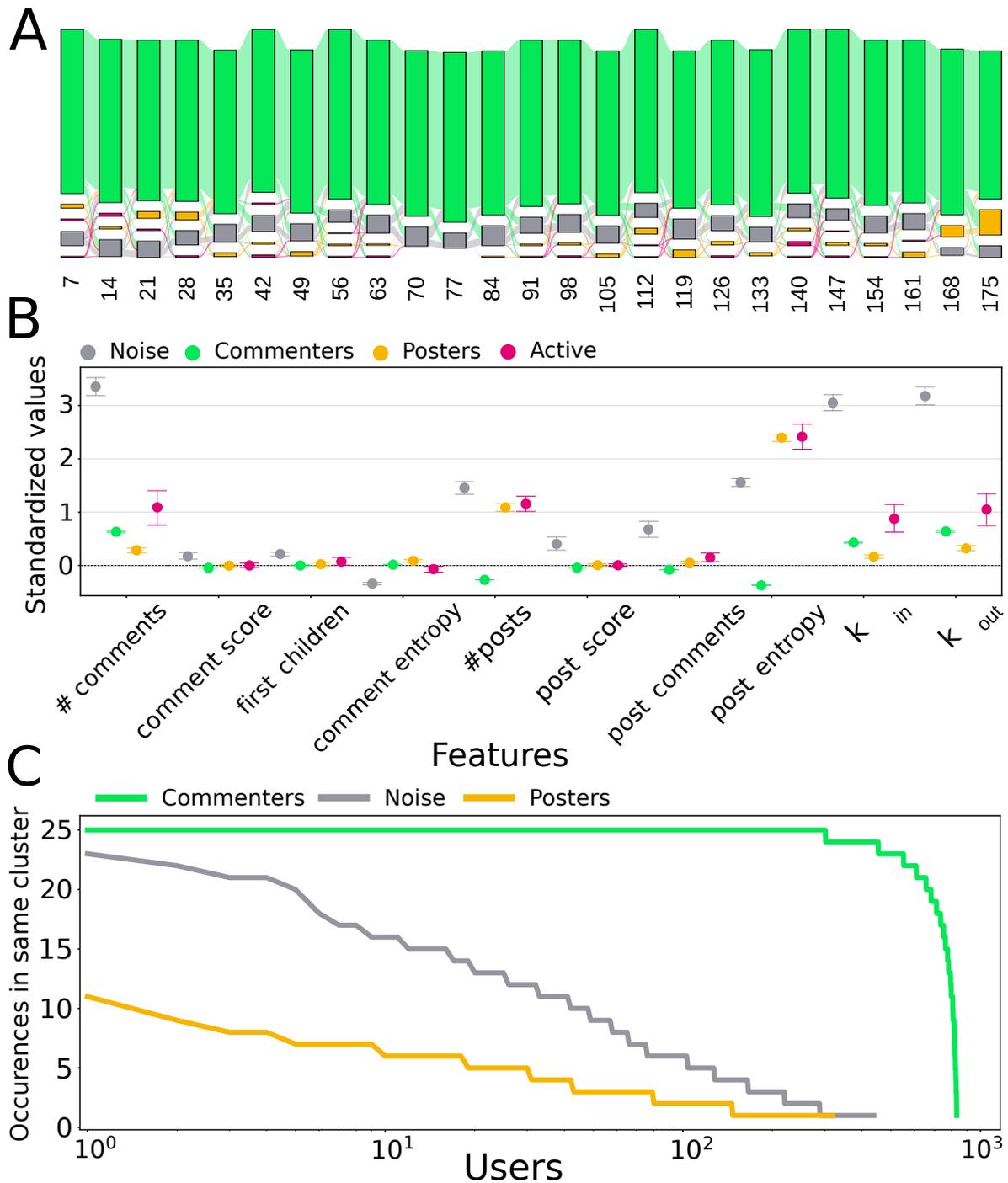}
    \caption{\textbf{Persistent users} A) Flow of users over the top three clusters and noise (in grey), we can see how most of the flux stays in the green cluster. B) Average values of the features for the 4 clusters, apart from noise we can identify the green cluster as commenters, the orange one as posters and the pink one as active users. C) Occurrences in the same cluster for all persistent users, we can see that most of them stay in the commenters cluster. }
    \label{fig_persistent}
\end{figure}

We select only users who are present in all non-overlapping weeks of the first part of our dataset, meaning all users who have been active at least once every week, leaving us with 832 individuals. In fact by selecting those persistent users we are able to follow their movements in time and see if they exhibit a certain trend, as is shown in the alluvial diagram of Figure \ref{fig_persistent} A. Figure \ref{fig_persistent} B shows the average feature values of each cluster, and helps us in identifying the largest cluster in green as that of commenters, the orange one of posters and the pink one of active users. Figure \ref{fig_persistent} C shows how many times a given user appears in the same cluster, highlighting that the majority of users remain in the commenters cluster, and a consistent fraction of users also appear in the noise cluster multiple times over the weeks. Instead we can see that the most persistent posters appear in the same cluster for less than half the total period of time, meaning that most of them are actually users from the other two clusters who on a given week decide to write more posts. However we still need to keep in mind that these users are only a small fraction of the total users of the community, and thus may not be able to capture the overall behavior.  
\newpage
\newpage
\section{Assortativity coefficient}\label{assortativity}
\begin{figure}[hbt!]
    \centering
    \includegraphics[width=0.8\textwidth]{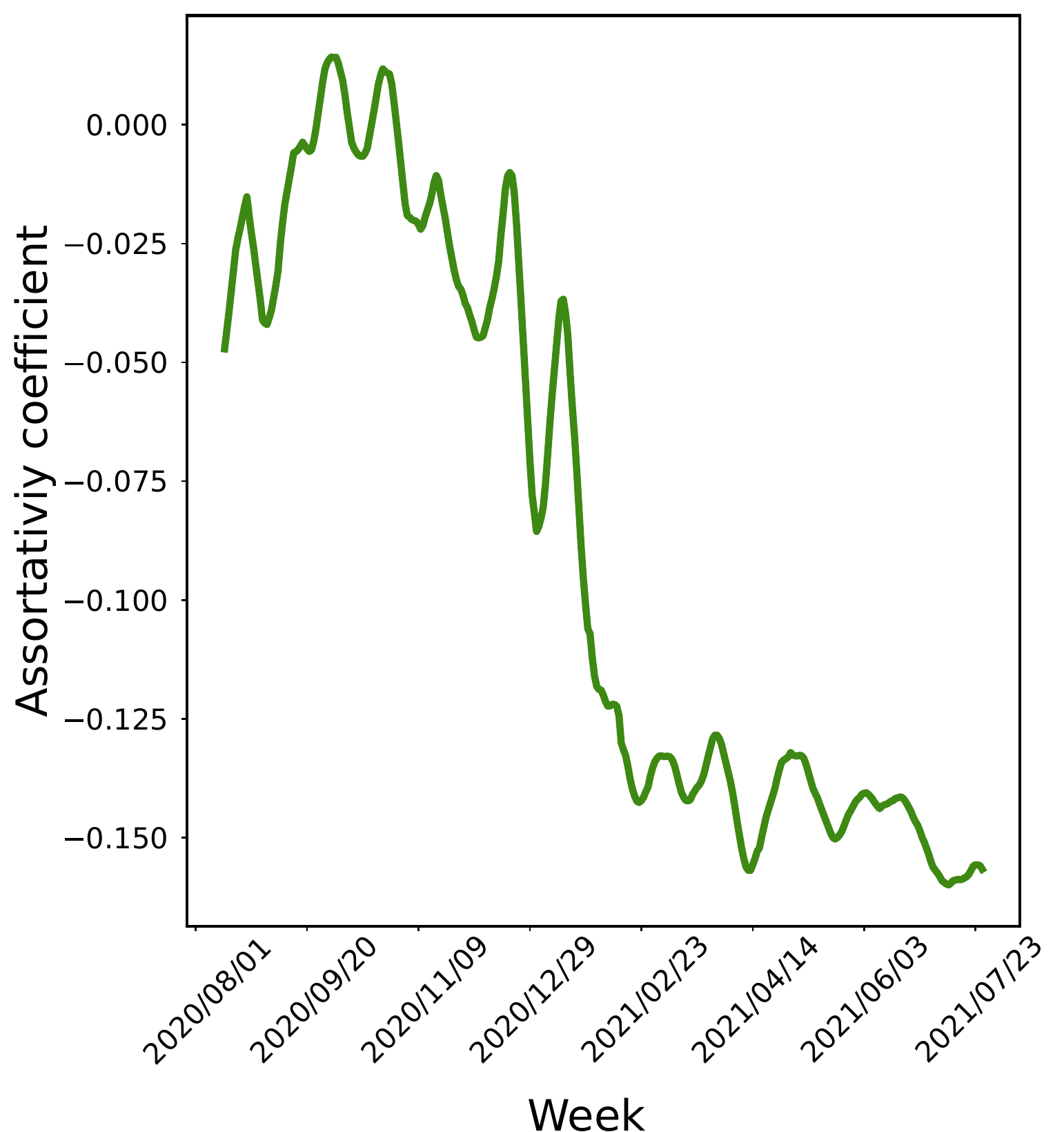}
    \caption{\textbf{Assortativity coefficient} Degree assortativity coefficient quantifying how the behavior of users after the short squeeze influences the network's topology. After January the networks become disassortative, where high degree users are connected to low degree ones.}
    \label{fig_assort}
\end{figure}

Assortativity in a network summarizes the tendency of nodes to connect to others who are similar to them, assortative mixing, or to those who are different, disassortative mixing \cite{Newman_2003}. In Figure \ref{fig_assort} we show how the degree assortativity coefficient of the network changes after the GameStop short squeese of January 2021. From values close to 0 pre-GME, to negative values from February onwards, this shift quantifies the different behavior of users in the subreddit. In fact conversation threads where users reply to each other via comments become less frequent, i.e. conversation trees have a low height and are made of mostly leaves (direct replies to the post). So the network becomes disassortative, and high degree nodes (users who write posts) are connected to low degree ones (users who write comments).

\end{document}